\pdfoutput=1

\documentclass{nature}

\usepackage{graphicx}

\usepackage{color}
\usepackage[normalem]{ulem} 
\usepackage{url}
\usepackage{amssymb,amsmath,amsfonts}		

\usepackage{pgffor}
\usepackage{tikz}
\usepackage[export]{adjustbox}

\newcommand{\e}{\mathrm{e}}

\newcommand{\E}[1]{{\mathrm E}\left[ #1 \right]}
\newcommand{\Var}[1]{{\mathrm Var}\left[ #1 \right]}
\newcommand{\Cov}[1]{{\mathrm Cov}\left[ #1 \right]}

\let\oldsqrt\sqrt
\def\sqrt{\mathpalette\DHLhksqrt}
\def\DHLhksqrt#1#2{%
\setbox0=\hbox{$#1\oldsqrt{#2\,}$}\dimen0=\ht0
\advance\dimen0-0.2\ht0
\setbox2=\hbox{\vrule height\ht0 depth -\dimen0}%
{\box0\lower0.4pt\box2}}

\title{Explaining the Prevalence, Scaling and Variance of Urban Phenomena}

\author{Andres Gomez-Lievano$^{*,1}$, Oscar Patterson-Lomba$^2$ \& Ricardo Hausmann$^{1,3,4}$}

\begin{document}

\maketitle

\begin{affiliations}
	\item Center for International Development, Harvard University, Cambridge, MA 02138, USA.
	\item Harvard T.H. Chan School of Public Health, Harvard University, Boston, MA 02115, USA.
	\item Santa Fe Institute, 1399 Hyde Park Road, Santa Fe, NM 87501, USA.
	\item Harvard Kennedy School, Harvard University, Cambridge, MA 02138, USA.
\end{affiliations}

\newpage

\begin{abstract}
The prevalence of many urban phenomena changes systematically with population size\cite{BettencourtPNAS2007}. We propose a theory that unifies models of economic complexity\cite{HidalgoHausmann2009,HausmannHidalgo2011} and cultural evolution\cite{Henrich2004Tasmanian} to derive urban scaling. The theory accounts for the difference in scaling exponents and average prevalence across phenomena, as well as the difference in the variance within phenomena across cities of similar size. The central ideas are that a number of necessary complementary factors must be simultaneously present for a phenomenon to occur, and that the diversity of factors is logarithmically related to population size. The model reveals that phenomena that require more factors will be less prevalent, scale more superlinearly and show larger variance across cities of similar size. The theory applies to data on education, employment, innovation, disease and crime, and it entails the ability to predict the prevalence of a phenomenon across cities, given information about the prevalence in a single city.
\end{abstract}

Scaling is ubiquitous across many phenomena\cite{Schroeder1991}, including physical\cite{Sornette2006} and biological\cite{WestBrown2005} systems, plus a wide range of human\cite{Gonzalez2008,McNerneyEtAl2011} and urban activities\cite{BettencourtPNAS2007,Batty2008}. Figure 1 shows, for US Metropolitan Statistical Areas, ten different phenomena classified in five broad types: employment, innovation, crime, educational attainment, and infectious disease. We observe scaling in the sense that the counts of people in each phenomenon scale as a power of population size. This relation takes the form $\mathrm{E}\{Y|N\}=Y_0~N^{\beta}$, where $\mathrm{E}\{\cdot|N\}$ is the expectation operator conditional on population size $N$, $Y$ is the random variable representing the output of a phenomenon in a city, $Y_0$ is a measure of general prevalence of the activity in the country, and $\beta$ is the scaling exponent, i.e., the relative rate of change of $Y$ with respect to $N$. From Fig. 1 we can also observe notable differences in the average prevalence, the slopes of the regression lines and the variance across all ten phenomena. Hence, we seek to explain four empirical facts: Prevalence follows a power-law scaling with population size, different phenomena have different general prevalence, different scaling exponents, and variance for cities of similar size. Remarkably, these observations appear to be pervasive across phenomena as we find them to be present in more than forty different urban activities. In this paper we propose a mechanism to explain them simultaneously.

\begin{figure}
		\centering
			\includegraphics[width=0.9\linewidth]{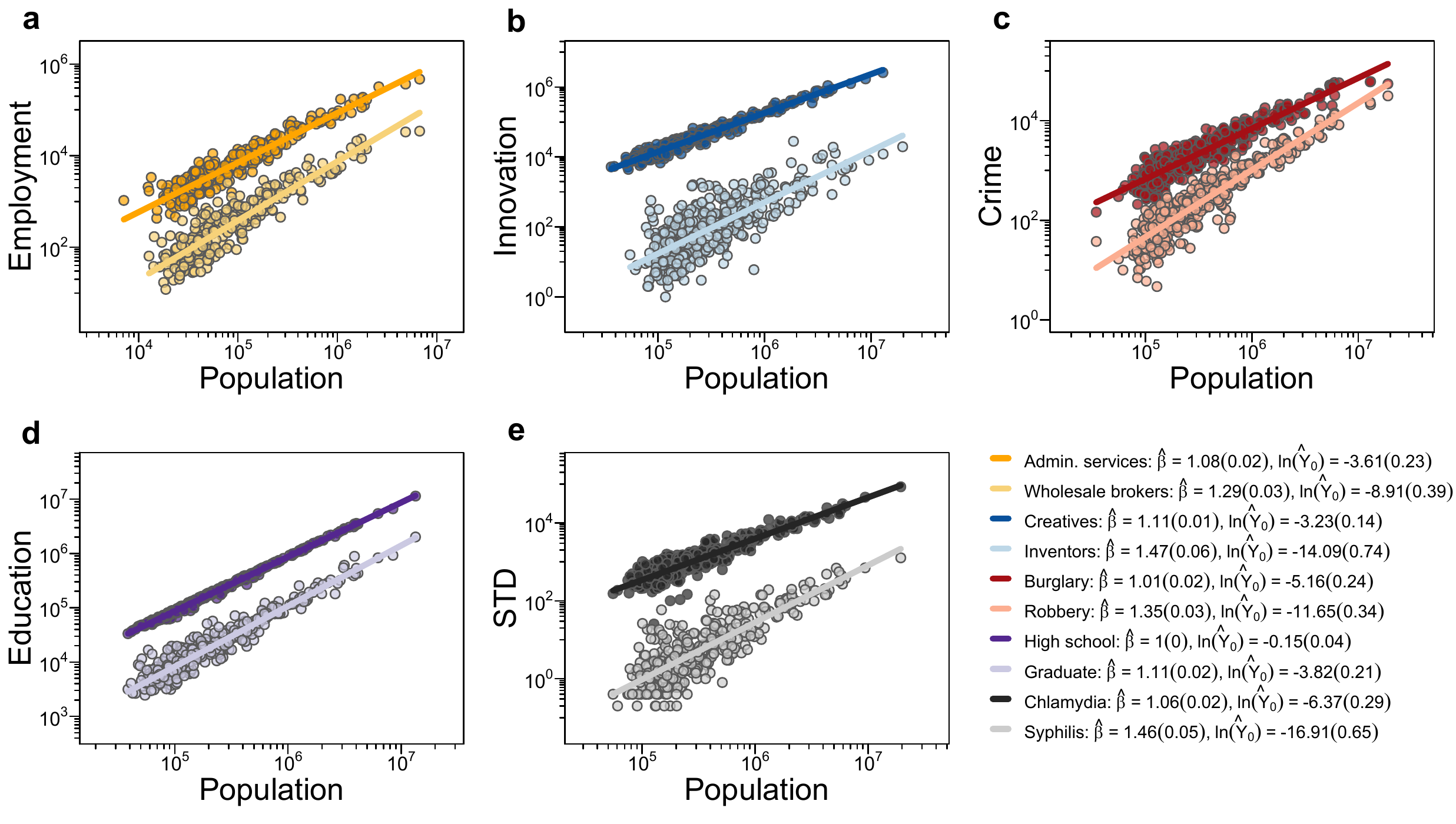}
		\caption{\footnotesize{Four facts across ten different urban phenomena we seek to explain. Prevalence follows a power-law scaling with population size, different phenomena have different general prevalence, different scaling exponents, and variance for cities of similar size. Cross-sections for (a) employment in two industries, (b) two types of innovative activities, (c) two types of violent crime, (d) people with a given educational level, and (e) two sexually transmitted diseases (see Methods section for data sources). The lines represent the best fit of the model $\mathrm{E}\{Y|N\}=Y_0 N^{\beta}$ (see Methods section for additional information).}}
\label{fig:crossSections}
\end{figure}

Scaling laws are important in science because they constrain the development of new theories: any theory that attempts to explain a phenomenon should be compatible with the empirical scaling relationships that the data exhibit. A number of mechanisms have been proposed to explain the origins of scaling. Most theories are based on a network description of the underlying phenomena and derive the scaling properties from the way the number of links grow with the number of nodes in the network, under some energy or budget constraints\cite{WestBrownEnquist1997,Banavar1999SizeAndForm,Arbesman2009,PanEtAl2013DensityDriven,Bettencourt2013,Yakubo2014PhysRevE}. Other scaling relationships are the result of how lines relate to surfaces, and surfaces to volumes\cite{McMahon1973,WestBrownEnquist1999,SamaniegoMoses2008,Banavar2010general}. We propose a different mechanism that improves on previous explanations in that it not only generates scaling, but also accounts for the value of the scaling exponent, the average relative prevalence across different phenomena, and the variance within phenomena across cities of similar size. 

The central assumption of our framework is that any phenomenon depends on a number of complementary factors that must come together for it to occur. More complex phenomena are those that require, on average, more complementary factors to be simultaneously present. This assumption is the conceptual basis for the theory of economic complexity\cite{HidalgoEtAl2007,HidalgoHausmann2009,HausmannHidalgo2011,KlimekHausmannThurner2012}. 

In addition, as with models of cultural evolution, we posit that the number of factors in the environment is a function of population size\cite{HenrichBoyd2002,Henrich2004Tasmanian,Powell2009}. Anthropological studies have shown this to be true about the diversity of skills, behaviors, beliefs, vocabulary and tools\cite{KlineBoyd2010,Mesoudi2011,Derex2013experimental,KempeMesoudi2014,CollardEtAl2013,Bromham2015}. More recent evidence of this relationship has been found in cities\cite{SFISummerSchool2012project,YounEtAl2015Universality,Bettencourt2014professional}. These models assume that cultural accumulation is a Darwinian process, in the sense that it involves inheritance, differential fitness and selection. The prediction is a logarithmic function of population size\cite{Henrich2004Tasmanian}. Our approach is not dependent on the precise justification for the logarithmic function,  since logarithms typically emerge from the fact that selection implies transforming initial distributions into extreme value distributions (such as a Gumbel distribution\cite{Henrich2004Tasmanian}) whose means grow logarithmically with sample size. For example, we can assume each factor has a different probability of appearance, and cities randomly sample from this distribution according to their size. If there is a process of selection, an extreme value distribution will emerge. In this setting, the diversity of factors will accumulate logarithmically with population size if the distribution of frequencies of the factors is Gumbel, meaning that the rarer factors will only appear in larger cities (see Supplementary Information for more details).

These two assumptions about complementarity and diversity are enough to generate our results. A wide range of phenomena including industrial employment, innovation, crime, educational attainment, and disease incidence are all statistically consistent with our theory. Moreover, we reveal an important empirical fact about the factors affecting different urban phenomena: that they change in similar ways across phenomena, implying that all scaling parameters for an urban phenomenon can be obtained from a single observation. This suggests that urban scaling is a highly constrained phenomenon, which in turn allows us to test the theory via its ability to predict the likely prevalence of a phenomenon across cities.

Our work is also related to the literature on production recipes\cite{AuerswaldEtAl2000}, which has been recently applied to explaining performance curves in production processes\cite{McNerneyEtAl2011}. The notion of complementarity, which is central in our approach, also plays a role in the ``Componential Theory of Creativity'' by T. Amabile\cite{Amabile1996creativity}, the ``violentization'' model of criminality of L.H. Athens\cite{Athens1992ViolenceSocialization}, and of recombinant growth models by S. Weitzman\cite{Weitzman1998}. The closest approach to our framework, however, is the model of Hausmann and Hidalgo\cite{HausmannHidalgo2011}, which assumes that industries are present in a location when the elements that are necessary for the industry are available in the location. They use a simple model in which the number of elements in a location is a binomial random variable with probability $r$ and the elements required by each industry is another binomial random variable with probability $q$. Assuming constant $r$ for all countries and $q$ for all industries they explain how ubiquitous industries are across countries, the inverse relationship between the diversity of countries and the average ubiquity of their industries, and other relevant statistics. However, they limit the analysis to industry presence and do not look at scaling phenomena. A novel conceptual component of our model is also to allow the required factors specific to a given activity to be different for each individual. That is, any two individuals in the population can require two different sets of factors in order to be counted into a given activity.

\begin{table}[tb]
	\caption{\footnotesize{Parameters of the model. The parameters $M$, $q$ and $r$ are in principle phenomenon-dependent.}}
	\label{tab:parameters}
	
	\small
	\centering
    \begin{tabular}{ll}
				\hline
    \textbf{Parameter}                                                      & \textbf{Meaning}                             \\ \hline
    $N>0$                                                  & City population size susceptible of participating of a given phenomenon.                           \\
    $M>0$                                                & Number of possible factors required for the given phenomenon.\\
    $q\in (0,1)$                                                            & Probability that an individual needs any given factor from the environment.\\
		$r\in (0,1)$                                                            & Probability that the city facilitates any one of the factors to the individual.\\\hline
    \end{tabular}
\end{table}

The parameters of the formal model are listed in Table 1. Each phenomenon has a number of factors $M$ on which it can depend. With probability $q$ an individual requires any one of those $M$ factors, and with probability $r$ a city provides any one of the factors. We model the random variable representing the aggregate output of a given phenomenon as $Y=\sum_{j=1}^N X_j$, where $X_j=1$ if individual $j$ has access to all the required factors she needs in city $c$ to be counted in a given activity, and $X_j=0$ if she does not, with $j\in\{ 1,\ldots, N \}$. 

Given a city with some factors present in it (from a total of $M$ possible factors), the probability that individual $j$ generates an output (i.e., that $X_j=1$), is the probability the individual requires none of the factors the city \emph{does not} have. Therefore, if an individual is exposed to $m$ factors, the individual cannot require any of the other $M - m$ factors that are not present, if his or her output is to be $1$. Since the probability that an individual does not require a particular factor is $1-q$, the probability that an individual is counted in the activity given a city with $m$ factors is $\Pr\{X_j=1|M_{\mathrm{city}}=m\}=(1-q)^{M -m}$, where $M_{\mathrm{city}}$ is a binomially distributed random variable $Binom(M, r)$.

It follows that $X_1,\ldots,X_N$ are identically distributed random variables. The expected value of $Y$ is thus $\mathrm{E}\{Y\}=N~\sum_{m=0}^{M} \Pr\{X_j=1|M_{\mathrm{city}}=m\}*\Pr\{M_{\mathrm{city}}=m\}$. The variance of $Y$ can be calculated similarly. This yields (see Supplementary Information for the complete derivation): 
\begin{equation}
	\mathrm{E}\{Y\}\approx N P,\label{eq:EYexp0}
\end{equation}
and
\begin{equation}
	\mathrm{Var}\{Y\}\approx \mathrm{E}\{Y\}^2\left(\frac{1}{\mathrm{E}\{Y\}} - \frac{1}{N} + \frac{1}{P^q}-1\right),\label{eq:varYbeforesigma0}
\end{equation}
where $P\equiv\mathrm{e}^{-M q(1-r)}$. 

Since $r$ is the fraction of factors an individual is expected to encounter in a city, $r$ represents a measure of urban diversity. This parameter captures the accumulation of factors in the population. As we have argued, factors tend to accumulate logarithmically with population size when a process of selection is involved (see Supplementary Information for more details). Factors can be acquired by individuals through a process of social learning as in models of cultural evolution, or by cities as a whole as they integrate individuals with qualitatively new and different characteristics, skills, behaviors, beliefs, occupations or tools. 

We thus assume that $r=a+b\ln(N)$. Replacing $r$ in Eq.\ref{eq:EYexp0} yields the scaling function $\mathrm{E}\{Y\}=Y_0~N^{\beta}$ (see Eqs.3 and 4 below). Hence, the power-law scaling of phenomena with population size across cities emerges from two relations that offset each other: the exponential relation between the prevalence of a phenomenon in a city and diversity, and the logarithmic relation of diversity with population size. We hypothesize that power-law scaling does not emerge if diversity does not scale logarithmically with population size. In this way, our theory can potentially reconcile observations in which power-law scaling breaks down (e.g., for small population sizes\cite{GomezLievanoYounBettencourt2012}), and can also be consistent with other scale-dependent functions, such as $\mathrm{E}\{Y\}=Y_0 N\ln(N/N_0)$ (see Refs. 39 and 40
), which can arise if diversity scales more slowly than logarithmically (see Ref. 33
). We thus provide theoretical support to a wide empirical literature on urban scaling\cite{BettencourtPNAS2007,GomezLievanoYounBettencourt2012,MantovaniRibeiroLenziPicoliMendes2013,ArcauteEtAl2014,Patterson2015STDs}. 

Furthermore, our model predicts that the logarithm of the general prevalence of a particular phenomenon, its scaling exponent, and the average standard deviation across population sizes, all change linearly according to the complexity of the phenomenon (see Supplementary Information for the precise derivation). Since the parameter $q$ is the fraction of factors an individual is expected to require from the city in order to be counted into a phenomenon, $q$ quantifies the complexity of that phenomenon. Specifically, we have
\begin{eqnarray}
	\ln(Y_0) &=& - M (1-a) q\label{eq:pred1},\\
	\beta-1  &=& M b q\label{eq:pred2},\\
	\sigma   &=& \sqrt{M (1-a-b \langle \ln N \rangle )}q\label{eq:pred3},
\end{eqnarray}
where $\sigma \equiv \sqrt{\langle \mathrm{Var}\{\ln Y\} \rangle}$, with $\langle \cdot \rangle$ being the mean across population sizes, such that $\langle \ln (N) \rangle$ is the mean of the logarithm of population sizes. In short, an increase in the complexity $q$ of a phenomenon (e.g., a decrease in transmissibility of a disease that makes it more difficult to acquire) would simultaneously decrease the intercept, increase the scaling exponent, and increase its variance in cities of same population size. In other words, complex phenomena are expected to be rare, scale steeply with population size, and their prevalence will be subject to high stochastic variability.

\begin{figure}[!ht]
		\centering
			\includegraphics[width=0.9\linewidth]{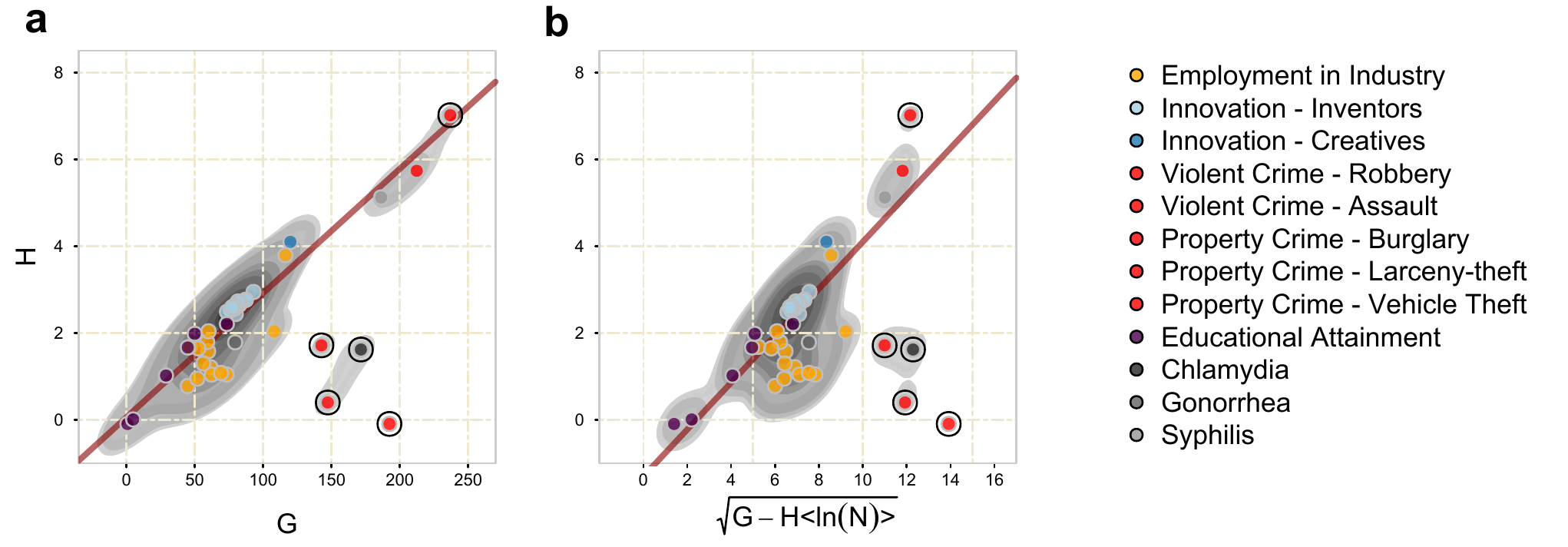}
		\caption{\footnotesize{Relationship between inferred values of parameters $G$, $H$ and $\sqrt{G-H \langle \ln(N)\rangle}$, across 43 different urban phenomena. The theory does not constrain their values, so the figure shows in gray the contours of a kernel-density estimate to reveal underlying patterns and relationships. A linear relationship is suggested by the estimated density. The line is the estimated robust regression that excludes the top 5 outliers marked with a solid circle which are phenomena with the least estimated density. In both panels the outliers are same: ``Robbery'', ``Aggravated Assault'', ``Burglary'', ``Larceny-theft'', and ``Chlamydia''. The linear trends in both panels are an empirical indication that the coefficients $s_1$ and $s_2$ are mostly constant across phenomena. See Methods section for more details. 
		}}
\label{fig:GandH}
\end{figure}
Conditioned on knowing $\beta$, $\ln(Y_0)$, and $\sigma$, Eqs.\ref{eq:pred1}, \ref{eq:pred2} and \ref{eq:pred3} represent three equations with four unknowns. The equations can then be solved for $G=M (1-a)$, $H=M b$, and $q$ (leaving $M$, the total possible number of factors that affect each phenomenon, undetermined). 

\begin{figure}[!ht]
		\centering
			\includegraphics[width=0.6\linewidth]{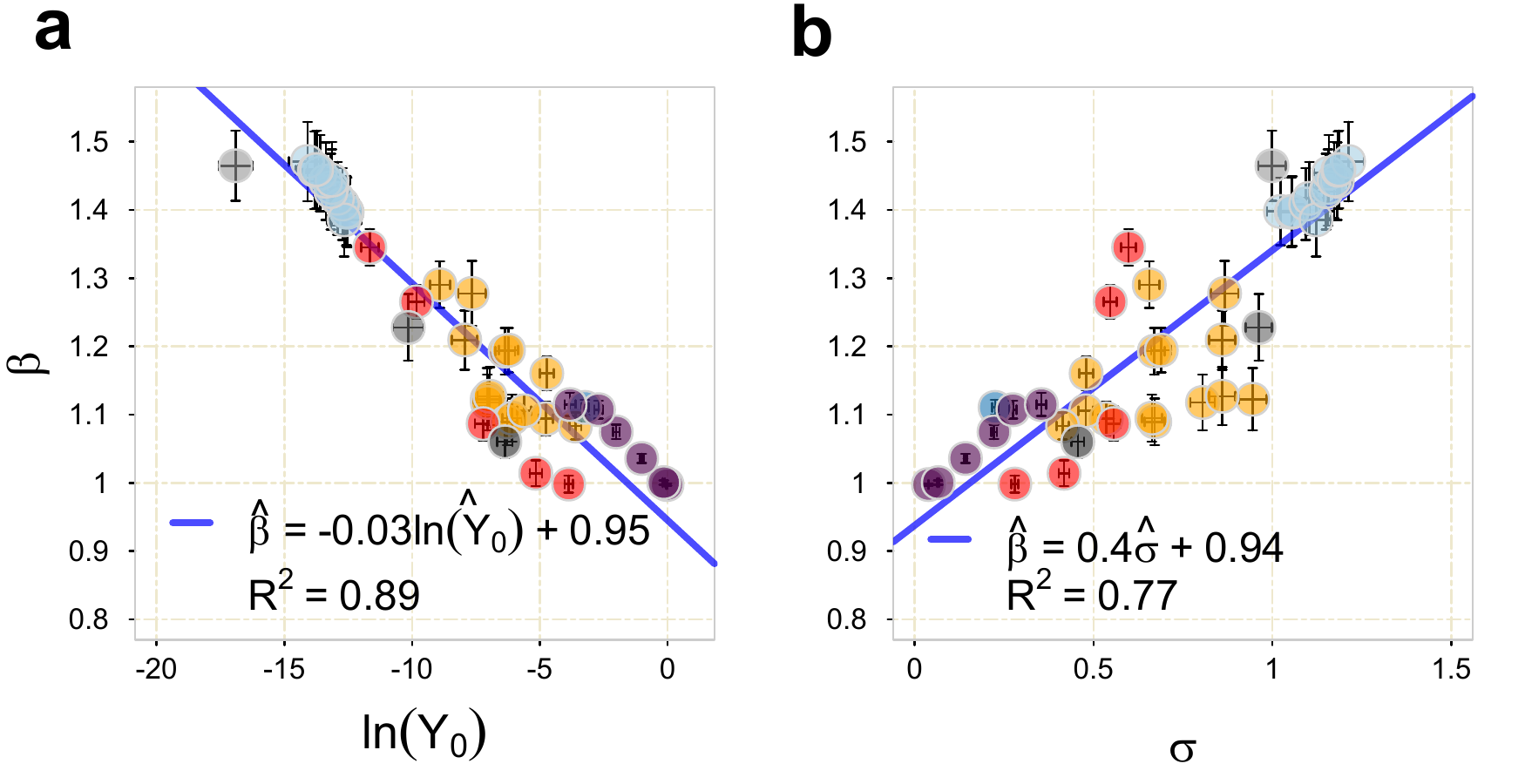}
		\caption{\footnotesize{The theory predicts a negative linear relationship between $\beta$ and $\ln(Y_0)$ (see panel {\bfseries a}), and a positive relationship between $\beta$ and $\sigma$ (see panel {\bfseries b}), both with an intercept of 1. As a consequence, there is an implied negative linear relationship between $\sigma$ and $\ln(Y_0)$ with no intercept. Both figures show the point estimates and the corresponding standard errors of the parameters of the scaling laws for each of the 43 urban phenomena studied. See Methods section for more details.
		}}
\label{fig:lnY0deltas}
\end{figure}
We estimate $\beta$ and $\ln(Y_0)$ through Ordinary Least Squares (OLS), and estimate $\sigma$ as the square root of the mean squared error of the OLS regression, and then solve for $G$, $H$ and $q$. Interestingly, even though $G$ and $H$ vary widely across phenomena, the ratio $s_1 = H/G$ remains numerically stable, as manifested in panel (a) of Figure \ref{fig:GandH} where $G$ and $H$ feature a linear relationship with no intercept. In this ratio the parameter $M$ factors out of $H$ and $G$ and cancels, yielding $s_1=b/(1-a)$. This suggests that the parameters for how diversity changes with population size (i.e., $a$ and $b$) are related in the same way across all phenomena. Similarly, the fact that $G$ is almost two orders of magnitude larger than $H$ signifies that the ratio $s_2 = H/\sqrt{G-H\langle\ln(N)\rangle}$ also remains approximately stable (panel (b) in Fig. \ref{fig:GandH}). This is because the ratio goes like $c \sqrt{G}$ with $c\rightarrow 0$. These ratios are important because they connect the scaling parameters. Namely, $\beta=1-s_1 \ln(Y_0)$ from Eqs. \ref{eq:pred1} and \ref{eq:pred2}, and $\beta=1+s_2\ \sigma$ from Eqs. \ref{eq:pred2} and \ref{eq:pred3}. As a consequence, the way $\beta$ changes with a change in $\ln(Y_0)$ and $\sigma$, respectively, is similar across activities. In other words, the implication of Fig. \ref{fig:GandH} is that we can plot the estimated values of $\beta$ vs. $\ln(Y_0)$ and $\beta$ vs. $\sigma$ for different activities in the same graph, and expect them to be linearly related. Figure \ref{fig:lnY0deltas} shows this is indeed the case. The implication is that these three scaling parameters are strongly constrained in the parameter space and lie in a line. 

\begin{figure}[!ht]
		\centering
			\includegraphics[width=\linewidth]{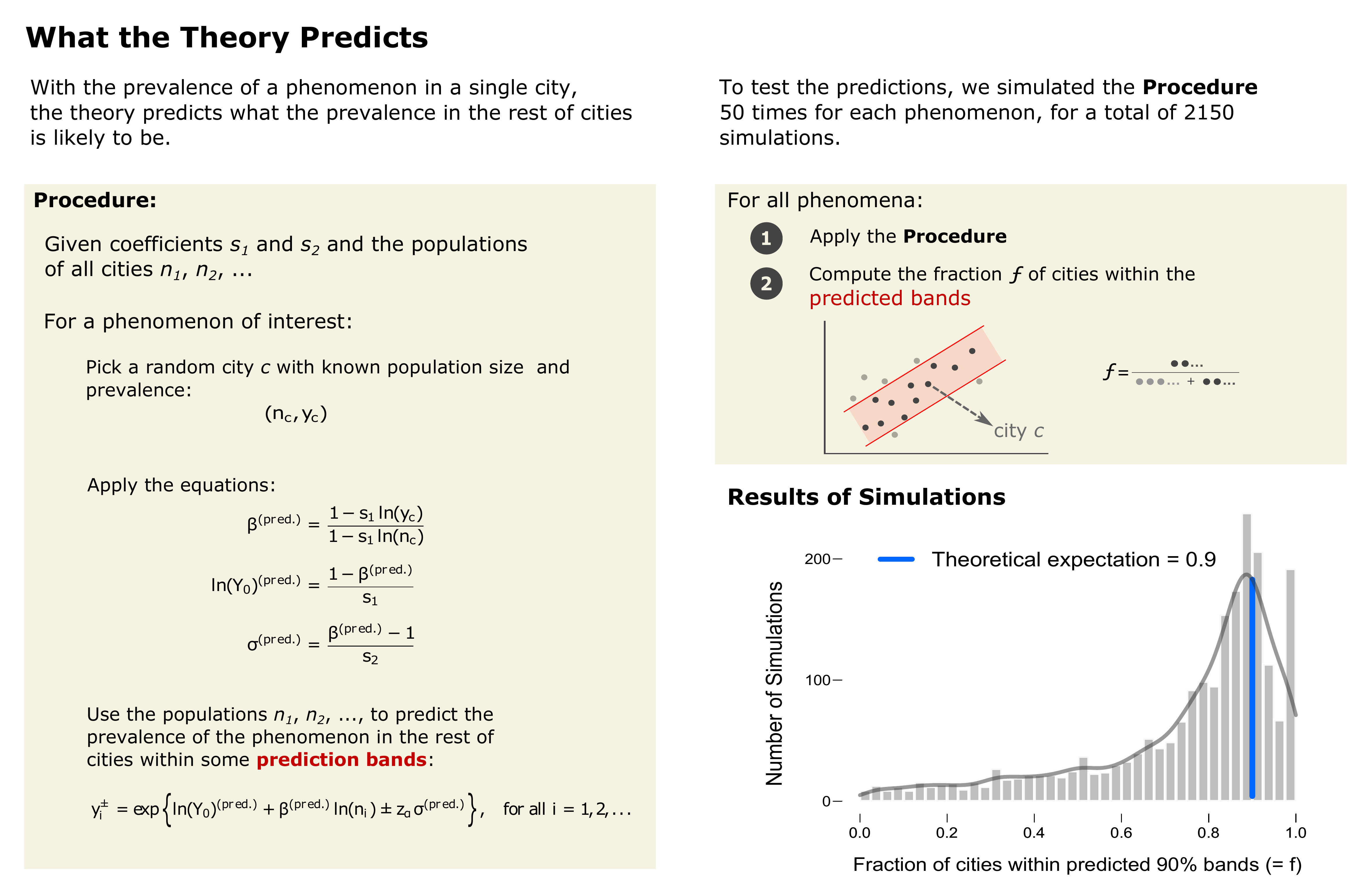}			
		\caption{\footnotesize{Testing the theory via its predictions. Knowledge of the prevalence and population size of a city determines the values of all the scaling parameters. The histogram comes from simulating the prediction procedure 50 times for each of the 43 urban phenomena in our database and computing the fraction of cities that fall within our predictions.
		}}
\label{fig:predprocedure}
\end{figure}

Provided the coefficients $s_1$ and $s_2$ are constants and are known in advance, the theory therefore establishes that knowing the value of one of the scaling parameters of a phenomenon of interest (exponent, general prevalence, or variance) determines the value of the others. If unknown, however, this one degree of freedom, in turn, can be fixed if we know the population $N=n_c$ and prevalence $Y=y_c$ in a single city $c$. This is possible if we assume the city is an average city, and the prevalence of the phenomenon is what is expected from its population size, $y_c = Y_0~n_c^\beta$. Thus, we can test the theory according to its ability to predict the prevalence of a phenomenon in other cities having knowledge of only one random data point (the prevalence of the phenomenon in a single city). Figure \ref{fig:predprocedure} explains the step-by-step procedure to determine bands between which the prevalence of a phenomenon is predicted to lie. To empirically test this, we use as an approximation the median of $s_1$ and $s_2$ across phenomena in our dataset, $s_1\approx 0.03045$ and $s_2\approx 0.33450$. We pick bands that are $z_{0.95}\approx 1.645$ standard deviations from the mean, so that if the theory is correct, 90\% of cities are expected to fall within the bands. For each of the 43 activities in our dataset, we simulated the procedure 50 times, picking a city at random each time (with replacement). The histogram of Fig. \ref{fig:predprocedure} shows the distribution of the fraction of cities $f$ that fell within the bands as a result of the $43\times50 = 2150$ simulations.

Here, we are using the proposed prediction framework to test the validity and scope of the theory. However, using this framework as an actual tool for predicting the prevalence of a phenomenon in cities where data is unreliable or unavailable is still premature. Further investigations and more data are needed to improve our theory and its practical utility. Moreover, it is important to keep in mind that our results so far imply that more complex phenomena have a higher variability. So even if the theory stands correct, 90\% prediction bands for complex phenomena can be as wide as two orders of magnitude and this intrinsic variability impacts on the practical use of such predictions.

There are two main reasons why some phenomena may deviate from our predictions. First, some of the counts for $Y$ are actually counts over a time period, which may shift arbitrarily the values that $\ln(Y_0)$ takes depending on the length of the period. For example, there is no reason why output must be computed as counts \emph{per year}, as opposed to \emph{per month}, or something else depending on the activity. And second, the scaling of output, according to the theory, is with respect to the \emph{potential} population $N$ which is ``susceptible'' of engaging in the activity or phenomenon (e.g., women, adults, or the working age population). Hence, $N$ is not necessarily the whole population of the city, and our estimations of $\ln(Y_0)$ carry that error from measuring incorrectly the size of the adequate population group. In spite of these effects the results in Fig. \ref{fig:lnY0deltas} are broadly consistent with the model. 

The theory we present is unabashedly simplified, avoiding issues about supply or demand, equilibrium, or the structure of social networks. We have assumed, for example, that people interact with the city as a whole, abstracting away interactions between individuals. We modeled each city as a set of factors, but we did not actually specify how factors appear. We introduced the notion of the complexity of a phenomenon, representing an average measure of how many inputs individuals need from the city to be able to be counted, or engage, in the given phenomenon. In the context of epidemiology, we have assumed the diversity of factors necessary for disease transmission to be mostly affected by socio-economic aspects, themselves subject to cultural evolution. Similarly with crime. Disease and crime, however, are the subject of strong public policy interventions aimed at reducing their influencing factors. How our model applies to these phenomena is a question that needs to be further analyzed as more data is collected.

We have also abstracted away important aspects of cities. First and foremost, we have presented a static view of cities. Also, we have bypassed the interdependencies between cities, and between activities, that arise from people migrating in and out of them\cite{Neffke2013SkillRelatedness}. Labor migration and the sharing of resources among cities in a region can affect the diversity of factors a city is exposed, and has access, to. Hence, factors imported from a wider region can affect the prevalence of urban phenomena. Further work is needed regarding the inclusion of these interactions into the model and their consequence on scaling. We have also left out the dynamic component involved when economic actors act according to complex decision rules. Finally, we have not taken into account the fact that economic and social actors exist not only at the level of individuals, but also at intermediate levels of organization such as families, neighborhoods, firms, and so on.

Accordingly, we do not expect predictions of this model to be numerically accurate, and yet they are quite reasonable. It is surprising that such a simple model can explain scaling, prevalence, and variance of such heterogeneous phenomena in an integrated framework. This indicates that the theory has captured something fundamental about social systems: namely, that they are complex, stochastic processes that involve many complementary factors accumulating through evolutionary processes. Thus, models that incorporate these elements can have broad applications in social science.

\begin{methods}

\subsection{Regression analysis.}
While our response variables $Y$ are conceptually counts, in practice, some of our data represent time averages or estimates from statistical offices. Still, we are trying to analyze under a unified framework our data which include both continuous and count variables. For count variables, the use of negative binomial, poisson, or zero-inflated regression analyses are preferred over ordinary least squares (OLS), given that the latter assumes a continuous normal conditional distribution of the response and does not allow for the use of zero counts when the regression is done over the logarithm of $Y$. All these methods should in principle yield similar coefficient estimates, and are rather intended to get better estimates of their standard errors. 

Since our analysis depends on comparing the estimated regression parameters across several urban phenomena, we have opted for the use of OLS regression for all phenomena throughout our analysis. The use of different regression models do not change dramatically our estimations, as expected.

\subsection{Kernel density estimation.}
In Figure 2 of the main text we show the values of $G$, $H$ and $\sqrt{G-H \langle \ln(N)\rangle}$, across 43 different urban phenomena. To reveal patterns in the distributions of these values we applied a 2-dimensional kernel density estimation separately for $G$ and $H$, and $\sqrt{G-H \langle \ln(N)\rangle}$ and $H$. See the Supplementary Information for an analysis of the outliers and how they affect the linear relationship.

We used the \textbf{R} package ``ks'', freely available on the web \cite{Duong2007}, which uses standard normal kernels with a conventional plug-in selector for the matrix bandwidth estimation. A useful feature of this package is that it allows non-zero values for the non-diagonal elements in the matrix.

\subsection{Data availability.}
The data sources are explained below. They have been aggregated and separated into different files which we provide in a ZIP file called ``Supplementary Data.zip'' that contains a single file for each urban phenomenon we studied (except for Sexually Transmitted Diseases that we kept in a single file), a README file, and a file ``ListUrbanPhenomena.xlsx'', which lists the different phenomena we used in our analysis with other parameters and field descriptions.

\begin{description}
	\item[Employees by industry.] 
	Data was downloaded using the programming codes that have made available by the Bureau of Labor Statistics through the website \url{http://www.bls.gov/cew/doc/access/data_access_examples.htm}. The specific data for micropolitan and metropolitan areas was selected using the guide in \url{http://www.bls.gov/cew/doc/titles/area/area_titles.htm}.

The metropolitan codes, however, are from the 2004 definitions. In \url{http://www.bls.gov/cew/cewfaq.htm#Q18}, it says
\begin{quotation}
	QCEW data for Metropolitan Statistical Areas (MSAs) for the years 1990 to present are based on the March 2004 MSA definitions. Aside from a few titling changes, there have been relatively few updates to those definitions since the March 2004 release. The next major revision to MSA definitions is expected in 2013. The QCEW program will release data for 2013 and forward based on those definitions.
\end{quotation}

However, these definitions do not match completely. From \url{http://www.bls.gov/news.release/metro.nr0.htm},
\begin{quotation}
	The Metropolitan New England City and Town Areas (NECTAs) and NECTA Divisions again are used for the six New England states, rather than the county-based delineations, for purposes of this news release.
\end{quotation}

The list of industry codes can be found in \url{http://www.bls.gov/cew/doc/titles/industry/industry_titles.htm}. We use employment numbers aggregated to 3-digit level industries. From the 91 different industries, we pick only those industries that have presence (at least 1 employee) in more than 250 metropolitan areas. This is to ensure the statistical significance is comparable with the other urban phenomena. Since our theory does not account for sublinear phenomena yet, we pick the industries that have scaling exponents of employment with population size larger than 1. This reduces the sample of 3-digit industries from 91 to 14. Our results, however, are robust to including more (superlinear) industries with presence in less than 250 MSAs.

	\item[Sexually transmitted diseases.]
	The data on Sexually Transmitted Diseases (STDs) consist of new cases of chlamydia and syphilis (primary, secondary and congenital). They represent the 5-year cumulative incidence, from 2007 to 2011, in the counties of the 48 contiguous states of the United States, as reported by the Centers for Disease Control and Prevention (CDC) \cite{cdcSTD12}. In our analysis we used the average of counts over the years 2007-2011.

The surveillance information in this dataset is based on the following sources of data: (1) notifiable disease reporting from state and local STD programs; (2) projects that monitor STD positivity and prevalence in various settings, including the National Job Training Program, the STD Surveillance Network, and the Gonococcal Isolate Surveillance Project; and (3) other national surveys implemented by federal and private organizations. This dataset does not include any individual-level information on reported cases. 

Since the STD data was originally obtained at the county level, we constructed MSA-level metrics using county-level data. See \cite{Patterson2015STDs} for details. Of the 375 MSAs within the 48 contiguous states, our dataset has information on 364.

	\item[Creative individuals.]
Here we use the definition of `creative occupations' given by the U.S. Department of Agriculture (USDA, \url{http://www.ers.usda.gov/data-products/creative-class-county-codes/documentation}), as an improvement to the originally proposed by Richard Florida \cite{Florida2004}. The USDA defines these occupations:
\begin{quotation}
	O*NET, a Bureau of Labor Statistics data set that describes the skills generally used in occupations, was used to identify occupations that involve a high level of ``thinking creatively.'' This skill element is defined as ``developing, designing, or creating new applications, ideas, relationships, systems, or products, including artistic contributions.''
\end{quotation}

The data is available at the county level and have to be aggregated using the 2003 MSA definitions which can be found at \url{http://www.census.gov/population/estimates/metro-city/0312msa.txt}. The number of MSAs according to this definition is 361 for the 48 contiguous states. To get the MSA populations we reconstruct it from Census tracks data, aggregating the 2010 populations of counties available at \url{https://www.census.gov/population/metro/data/c2010sr-01patterns.html}. 

	\item[Inventors.]
Counts of inventors are publicly available through the U.S. Patent and Trademark Office website at \url{http://www.uspto.gov/web/offices/ac/ido/oeip/taf/inv_countyall/usa_invcounty_gd.htm}. According to the link (\url{http://www.uspto.gov/web/offices/ac/ido/oeip/taf/reports.htm}) ``[t]his report applies to U.S. resident inventors who have received a utility patent (i.e., ``patent for invention'') granted by USPTO since 2000. The report includes a series of tables that display U.S. states and the regional components (e.g., counties) in which the inventors resided. Counts of the inventors and their patents are provided for each of the regional components.''

The documentation can be found in \url{http://www.uspto.gov/web/offices/ac/ido/oeip/taf/inv_countyall/usa_invcounty_gd.htm}. In Figure 2 and 3, we plotted the years 2000 to 2013 using the 2013 definition of Metropolitan Statistical Areas in terms of counties according to the U.S. Census Bureau (see \url{https://www.census.gov/popest/data/metro/totals/2013/CBSA-EST2013-alldata.html}). We merged to this dataset the MSA populations, from 2000 to 2013, reported by the Bureau of Economic Analysis.

	\item[Crime.]
Data for different types of crimes at the MSA level is collected by the Federal Bureau of Investigation (FBI). These data is publicly available at official the website \url{https://www.fbi.gov/about-us/cjis/ucr/crime-in-the-u.s/} for different years. In our study, we limited our analysis to the years 2010, 2011 and 2012.

Two important caveats about the crime statistics that we used in our analysis are in place. On the one hand, we would ideally like to have the counts over some period of time of unique individuals that were victims of different types of crimes (we would also like to have counts of criminals in urban areas, but this is obviously data difficult to measure). We have proxied the number of victims by the counts of crimes. On the other hand, our model provides predictions for counts of people $Y$ that engage in a given activity, and we compare these counts with the population $N$ that is \emph{susceptible} to this activity. For most activities $N$ is easy to define and is typically the total population size of a city. For other activities, $N$ is not so easy to define. Hence, we have removed from our analysis (see Figure 2 and 3) ``murder and nonnegligent manslaughter'' and ``forcible rape''. The relevant population $N$ that corresponds to these types of violent crimes is not the total population size of a city, it represents a restricted part of the total population, and we think these phenomena require analysis that is out of the scope of our model. For instance, forcible rape, as defined by the FBI (see \url{https://www.fbi.gov/about-us/cjis/ucr/crime-in-the-u.s/2010/crime-in-the-u.s.-2010/violent-crime/rapemain}) is ``the carnal knowledge of a female forcibly and against her will''. Misspecifications of $N$ in our regressions produce a bias in the estimation of $\ln(Y_0)$. We avoid such misspecifications by removing these two types of crimes from our analysis.

	\item[Educational attainment.]
We have used the estimates of the population by the different types of educational attainment from the 2009-2013, 5-Year American Community Survey (ACS) from the U.S. Census Bureau. We have used as the base population $N$ the population of 25 years and older. 

This dataset is accessible through the website American FactFinder, at \url{http://factfinder.census.gov/}. Selecting Advanced Search, entering ``S1501'' as the topic, corresponding to Educational Attainment. We selected the 2009-2013, 5-Year ACS data, and in ``Geographies'' we selected data for all U.S. Metropolitan Statistical Areas. 

We adjusted the educational attainment categories to reflect increases in complexity. Hence, from least to most complex, we defined six activities: (1) 9th grade, or higher, (2) High school graduate, or higher, (3) Some college, or higher, (4) Associate's degree, or higher, (5) Bachelor's degree, or higher, and (6) Graduate or professional degree.
\end{description}

\end{methods}



\begin{description}
 \item[Acknowledgments.] We thank A.-L. Barabasi, J. Lobo, L.M.A. Bettencourt, F. Neffke, S. Valverde, D. Diodato and C. Brummitt for their useful comments on this work. We also thank M. Akmanalp and W. Strimling for their suggestions about esthetics. This work was funded by the MasterCard Center for Inclusive Growth, and Alejandro Santo Domingo. O.P-L. acknowledges support by National Institutes of Health (NIH) grant T32AI007358-26. 
	\item[Author contributions.] A.G-L. and O.P-L. collected the data, conceived and designed the study. A.G-L. conducted the analyses. A.G-L. and R.H. developed the model. A.G-L., O.P-L. and R.H. wrote the manuscript. All three authors reviewed and approved the paper.
	\item[Competing interests.] The authors declare no competing interests. 
	\item[Materials \& Correspondence.] The sources of data used in this paper are available to the public and are reported in the Methods and Supplementary Information. Correspondence and requests for materials should be addressed to A.G-L.~(email: Andres\_Gomez@hks.harvard.edu).
\end{description}


\newpage
\appendix 
\begin{flushleft}
{\Large
\textbf{Supplementary Information:\\
Explaining the prevalence, scaling and variance of urban phenomena}
}
\\
\vskip 0.5 cm
Andres Gomez-Lievano$^{1,\ast}$, 
Oscar Patterson-Lomba$^{2}$,
Ricardo Hausmann$^{1,3,4}$
\\
\bf{1} Center for International Development, Harvard University, Cambridge, MA.
\\
\bf{2} Harvard T.H. Chan School of Public Health, Harvard University, Boston, MA.
\\
\bf{3} Santa Fe Institute, 1399 Hyde Park Road, Santa Fe, NM.
\\
\bf{4} Harvard Kennedy School, Harvard University, Cambridge, MA.
\\
$\ast$ E-mail: Corresponding andres\_gomez@hks.harvard.edu

\end{flushleft}

\makeatletter
\renewcommand{\fnum@figure}{Supplementary Figure \thefigure}
\makeatother

\renewcommand\refname{Supplementary References}

\section{Supplementary Discussion}
\subsection{Toy Example for How Complementarity Works}
\begin{figure}[!ht]
		\centering
			\includegraphics[width=0.6\textwidth]{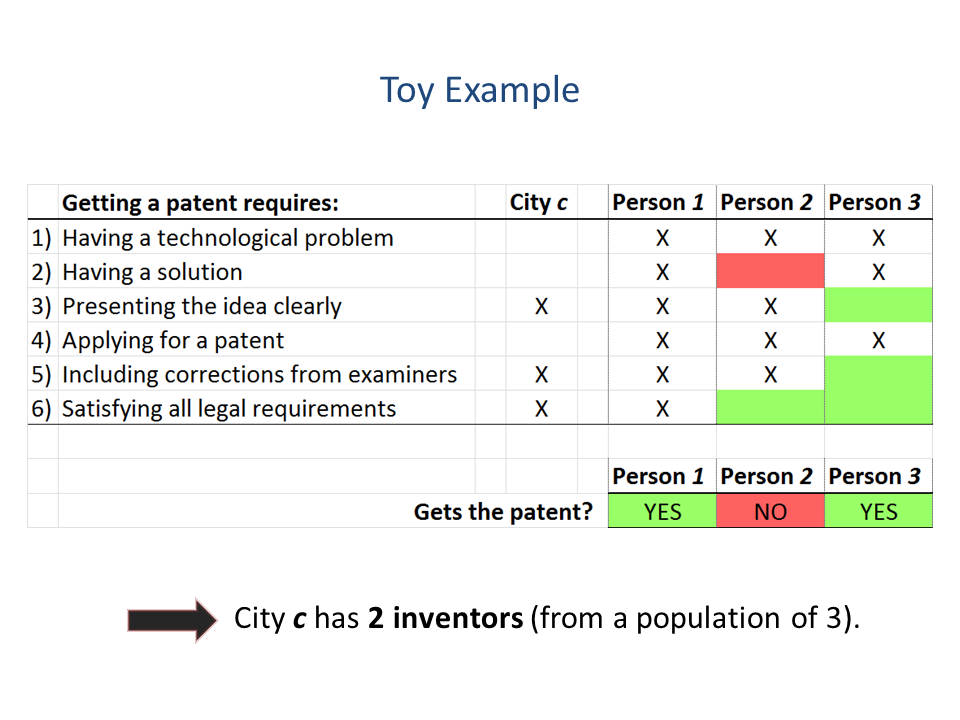}
		\caption{Toy example of how the model works. A given activity, in this case inventive activity or patenting, requires a set of substeps to be counted in. Some of the substeps may be facilitated by the individual and some by the city. Only when all the substeps are satisfied the person is counted in the activity.}
\label{fig:toyexample}
\end{figure}

We introduce our model with a simplified example.\footnote{The scheme we present is inspired by \cite{Shockley1957}.} To get a patent, one must \emph{(i)} have a technological problem, \emph{(ii)} have a solution, \emph{(iii)} present the idea clearly, \emph{(iv)} apply for a patent, \emph{(v)} include subsequent corrections from examiners, and \emph{(vi)} satisfy all the legal requirements. Supplementary Figure \ref{fig:toyexample} is a schematic representation of this example. Analogous schemes apply to getting sick, a job, a degree, committing a crime, and other many activities. The complementarity principle (one could also refer to it as the Anna Karenina's Principle) establishes that if one or more of the requirements \emph{(i)-(vi)} is missing, the person fails to do the activity. In this example there are $2^6$ total possibilities, only one leads to successful output.

In principle, the larger the number of conditions, namely $M$, that must be met to get an outcome from any given activity (in the patent example before, $M=6$), the more difficult it is that the outcome will occur. In practice, however, rather than the number of conditions, the difficulty of an activity depends more on how easily are these conditions facilitated by the \emph{person} involved in the given activity or by her \emph{environment}. In the patent example, for instance, one person may be a lawyer and thus readily fulfill condition number \emph{(vi)}, while another may require it from an external source (by hiring a lawyer, for instance). It is then up to the environment whether the second person will be able to file the patent successfully or not (assuming both have satisfied substeps \emph{(i)}-\emph{(v)}). 

Figure \ref{fig:toyexample} gives the example of three individuals that live in a city. This particular city facilitates three of the six substeps required to become an inventor (marked with an \textbf{\textit{X}}). The first individual possesses all the elements to become an inventor (also marked with an \textbf{\textit{X}}). As a consequence, this person can become an inventor regardless of what city she finds herself in. The second person is missing the substeps \emph{(ii)} and \emph{(vi)}. At the end, this person does not patent because he is unable to fulfill substep \emph{(ii)}, because he lacks it and the city he lives in does not help him with it. The third person, in contrast, is able to get a patent because she fulfills all the requirements, even though she relies more on her environment than the second person.

\subsection{Solving the Model}
Before we start solving the model, let us recall our question of interest. Given that urban phenomena in general require the coordination of several factors, we want to understand how the inherent difficulty of a given urban activity interacts with the city population size, such that we observe the stylized facts discussed in the main text. 

The total output of the city is the sum of the output across all individuals in the city. Given the two sources of randomness in our model (the list of factors of the city and the diversity of requirements across individuals), we want to know the statistical characteristics of total output. The random variable representing the total output in a city is expressed as
\begin{equation}
	Y = \sum_{j=1}^{N} X_j,
\end{equation}
and we thus need to understand the statistics of individual output, $X_j$.

\subsubsection{The Probability of Being Counted in an Urban Phenomenon}
Given a city $c$ with $m$ factors present in it (from a total of $M $), the probability that individual $j$ generates an output (i.e., that $X_j=1$), is the probability the individual requires any number of the $m$ factors that the city has (from $0$ to $m$), but \emph{none} of the ones that the city \emph{does not} have. Therefore, if the city has $m$ factors, the individual cannot require any of the other $M -m$ factors if his or her output is to be $1$. The probability that the individual does not require a single factor is $1-q$. Hence, 
\begin{equation}
	\Pr\{X_j=1|M_{\text{city}}=m\}=(1-q)^{M -m},
\label{Aeq:condsingleoutput}
\end{equation}
where $M_{\text{city}}$ is binomial random variable with parameters $M$ and $r$.

Using Eq.\ref{Aeq:condsingleoutput} we can now answer what is the expected value of $Y$, conditioned on the city having $m$ factors, since $\E{X_j|M_{\text{city}}=m}=\Pr\{X_j=1|M_{\text{city}}=m\}=\e^{(M -m)\ln(1-q)}\approx \e^{-(M -m)q}$ (for small $q$). To get the expectation for the aggregate count we multiply by the total population. Using the approximation $\ln(1-\epsilon)\approx -\epsilon$ for $\epsilon\ll 1$, we get
\begin{equation}
	\E{Y|M_{\text{city}}=m}\approx N\e^{-(M -m)q}.
\label{Aeq:EYgivencity}
\end{equation}  
Hence, the number of factors $m$ that an individual encounters, and the complexity of the activity $q$, have opposing effects on the probability of engaging in the activity: the former increases the probability while the latter reduces it.

\subsubsection{Mean, Variance and Covariance of Binary Random Variable of Person}
Now, individuals are exposed to different numbers of factors in a city, but they are exposed \emph{on average} to $M r$ factors. Hence, we will use the law of total probabilities to sum over all possible $m$. One could expand the model further to include heterogeneity within the city. Hence, the number of factors an individual is exposed to would be modeled as $Binom(M,r(x,y))$, where $r(x,y)$ can vary within the city as a function of geographical coordinates $(x,y)$. In addition, each factor would be sampled differently, so that the sampled random factors individuals are exposed to, and/or require in their activities, are interdependent depending on the geographical (and cultural, professional, or ethnic) proximity\cite{LehmannAokiFeldman2011}. In what follows, however, we will assume $r$ is a constant throughout the city.

Since the variable $X_j$ is binary, there are simplifications that are possible when calculating expectations and variances. Hence,
\begin{align}
	\E{X_j}&=\Pr\{ X_j=1 \},
\end{align}
and
\begin{align}
	\Var{X_j}&=\Pr\{ X_j=1 \} - \Pr\{ X_j=1 \}^2.
	\label{eq:varX}
\end{align}

One can also calculate the covariance, such that:
\begin{align}
	\Cov{X_i, X_j} &= \E{X_i X_j} - \E{X_i}\E{X_j}, \nonumber\\
	&= \Pr\{ X_i=1, X_j=1 \}  - \Pr\{ X_i=1 \}\Pr\{ X_j=1 \}, \nonumber\\
	&= \Pr\{ X_i=1, X_j=1 \} - \Pr\{ X_j=1 \}^2 \label{Aeq:covXiXj}.
\end{align}
To compute $\E{X_j}$, $\Var{X_j}$, and $\Cov{X_i, X_j}$, we need to calculate $\Pr\{X_j=1\}$ and $\Pr\{X_i=1, X_j=1\}$. 

We will present two different ways of computing these probabilities. We present both ways for illustrative purposes, but also to check our results are correct.

To calculate $\Pr\{ X_j=1 \}$ we first condition on the person requiring $m$ factors, we calculate the probability that the city has those $m$ factors, and sum over all possible $m$:
\begin{align}
	\Pr\{ X_j=1 \} &= \nonumber\\
	\sum_{m=0}^{M } \Pr\{ X_j=&1 | M_{\text{person}} = m \}\Pr\{ M_{\text{person}}=m \}, \nonumber\\
	&=\sum_{m=0}^{M } r^m ~ \binom{M }{m}q^m(1-q)^{M -m}, \nonumber\\
	&=\left[ rq+1-q \right]^{M }, \nonumber\\
	&=\left[ (1-r)(1-q) + r \right]^{M }.\label{Aeq:Pj}
\end{align}

To calculate the joint probability, we use a similar method. We assume that the city has $m$ factors, and we use Eq.\ref{Aeq:condsingleoutput}, to add over all values of $m$. The main advantage of this second way of calculating probabilities is that by conditioning on the city having $m$ factors, we can use the fact that $X_i$ and $X_j$ become conditionally independent: 

\begin{align}
	\Pr\{ X_i=1, X_j=1 \} &= \sum_{m=0}^{M } \Pr\{ X_i=1, X_j=1 | M_{\text{city}} = m \}\Pr\{ M_{\text{city}}=m \}, \nonumber\\
	&= \sum_{m=0}^{M } (1-q)^{M -m} (1-q)^{M -m} \binom{M }{m}r^m(1-r)^{M -m}, \nonumber\\
	&= \left[ (1-r)(1-q)^2 + r \right]^{M }.\label{Aeq:joint}
\end{align}


Using Eqs.\ref{Aeq:Pj} and \ref{Aeq:joint} to calculate the covariance given by Eq.\ref{Aeq:covXiXj}:
\begin{align}
	\Cov{X_i, X_j}&=\left[ (1-r)(1-q)^2 + r \right]^{M } \nonumber\\
	& \quad \quad - \left[ (1-r)(1-q) + r \right]^{2M }.
\label{Aeq:indcovar}
\end{align}

Using the approximation whereby $\ln(1-\epsilon)\approx -\epsilon$, for $\epsilon\ll 1$, we can write Eq.\ref{Aeq:Pj} and Eq.\ref{Aeq:joint} as
\begin{equation}
	\Pr\{ X_j=1 \}\approx \e^{-M q(1-r)},
\end{equation}
and
\begin{align}
	\Pr\{ X_i=1, X_j=1 \}&\approx \e^{-M q(2-q)(1-r)},\nonumber\\
	&=\Pr\{ X_j=1 \}^{2-q}.\label{Aeq:approxPXiXj}
\end{align}

To simplify notation, from this point forward, let $P\equiv\Pr\{X_j=1\}$ be the marginal probability that $X_j=1$, which is independent of the person $j$, as shown by Eq.\ref{Aeq:Pj}.

\subsubsection{Mean and Variance of Total Output}
\label{sec:meanandvar}
We now calculate the mean and variance of $Y=\sum_{j=1}^N X_j$. For the mean we get
\begin{align}
	\E{Y}&=\sum_{j=1}^N\E{X_j},\nonumber\\
	&=N \left[ (1-r)(1-q) + r \right]^{M },\nonumber\\
	&\approx N \e^{-M q(1-r)},\label{Aeq:EYexp}
\end{align}
where we have used the approximation $P\approx\e^{-M q(1-r)}$, for $q(1-r)\ll 1$, in the last step. And for the variance we get
\begin{align}
	\Var{Y}&=\sum_{j=1}^N\Var{X_j} + \sum_{i\neq j} \Cov{X_i, X_j},\nonumber\\
	&=N\left[ \left[ (1-r)(1-q) + r \right]^{M }\right. \nonumber\\
	& \left. \quad \quad - \left[ (1-r)(1-q) + r \right]^{2M } \right]\nonumber\\
	&\quad + N(N-1)\left[\left[ (1-r)(1-q)^2 + r \right]^{M } \right.\nonumber\\
	&\left. \quad \quad - \left[ (1-r)(1-q) + r \right]^{2M }\right].\label{Aeq:fullVar}
\end{align}

Equation \ref{Aeq:fullVar} can be simplified by writing it in terms of $P$:
\begin{align}
	\Var{Y}&\approx NP(1-P) + N^2\left(P^{2-q}-P^2\right),
\end{align}
where we have used the approximation of Eq.\ref{Aeq:approxPXiXj}, and assumed $N$ is large enough so that $N-1\approx N$. Expanding and factoring out $N^2P^2$ yields
\begin{align}
	\Var{Y}&\approx (NP)^2\left(\frac{1}{NP} - \frac{1}{N} + \frac{1}{P^q}-1\right),\nonumber\\
	&=\E{Y}^2\left(\frac{1}{\E{Y}} - \frac{1}{N} + \frac{1}{P^q}-1\right).\label{Aeq:varYbeforesigma}
\end{align}

\subsubsection{Probability Distribution of Total Output}
We can also compute the probability distribution of $Y$ in a similar way as we did for the calculation of the joint probability $\Pr\{X_i, X_j\}$. That is, we first condition on the city having $m$ factors, and sum over the values of $m$:
\begin{align}
	\Pr\{ Y=k \} &= \sum_{m=0}^{M } \Pr\{ Y=k | M_{\text{city}}=m \}\Pr\{ M_{\text{city}}=m \},\nonumber\\
	&= \sum_{m=0}^{M } \binom{N}{k} \Pr\{ X_1=1,\ldots,X_k=1, \nonumber\\
	&\quad \quad X_{k+1}=0,\ldots,X_N=0 | M_{\text{city}}=m \}\Pr\{ M_{\text{city}}=m \},
\end{align}
where we are using the fact that the $X_i$'s are exchangeable (e.g., $\Pr\{ X_1=1, X_2=1, X_3=0 \}=\Pr\{ X_1=1, X_2=0, X_3=1 \}=\Pr\{ X_1=0, X_2=1, X_3=1 \}$), and thus we are counting all the ways in which $k$, out of the $N$ citizens, generate an output.

Recalling Eq.\ref{Aeq:condsingleoutput}, and the fact that individuals are conditionally independent, we get that
\begin{align}
	\Pr\{ Y=k \} &= \sum_{m=0}^{M } \binom{N}{k} \left[(1-q)^{M -m}\right]^{k} \left[1-(1-q)^{M -m}\right]^{N-k} \binom{M }{m}r^m(1-r)^{M -m}.
\label{Aeq:Yprob}
\end{align}

Depending on the values of the parameters, Eq.\ref{Aeq:Yprob} is a probability function that can generate skewed random variables. 

Since the output of individuals is positively correlated according to Eq.\ref{Aeq:indcovar}, the condition of independence in the Central Limit Theorem is violated. Hence, it is not surprising that $\Pr\{ Y=k \}$ does not approximate a normal distribution (or a binomial, if we keep $Y$ discrete). This is consistent with the fact that total output in cities has been found to be lognormally distributed \cite{BettencourtPLOS2010, GomezLievanoYounBettencourt2012, AlvesRibeiroMendes2013, AlvesRibeiroLenziMendes2013, AlvesRibeiroLenziMendes2014, MantovaniRibeiroLenziPicoliMendes2013}.

\subsubsection{How Complexity Affects Variance}
As explained above, the variance of $Y$ is given by
\begin{align}
	\Var{Y}&=\E{Y}^2\left(\frac{1}{\E{Y}} - \frac{1}{N} + \frac{1}{P^q}-1\right).\label{eq:varYbeforesigma}
\end{align}
Notice that $\Var{Y}$ and $\E{Y}$ are functions of population size $N$ that can also be though of as a random variable. This is important, since in the text we assume the parameter $\sigma$ to be a measure of (root square) variance \emph{averaged} over population sizes.

Let us write Eq.\ref{eq:varYbeforesigma} as
\begin{align}
	\Var{Y}&=\E{Y}^2\left(\e^{\sigma^2(N)}-1\right)\label{eq:varYaftersigma},
\end{align}
where we have defined the function $\sigma^2(N)$ as
\begin{align}
	\sigma^2(N)&=\ln\left(\frac{1}{\E{Y}} - \frac{1}{N} + \frac{1}{P^q}\right) \label{Aeq:sigma2func}.
\end{align}

Since the probability function of $Y$ is approximately a lognormal distribution, the function $\sigma^2(N)$ represents the average variance of the logarithm of $Y$:
\begin{equation}
	\sigma^2(N)\approx \Var{\ln(Y)}.
\end{equation}
Therefore, $\sigma^2(N)$ represents the variance in the vertical direction (for a given $N$) in any of the cross sections shown in Figure 1 of main text. This is because the plots are logarithmic scales, such that what we see is not the spread of $Y$, but the spread of $\ln(Y)$. 

Assuming $N\gg 1$, then
\begin{align}
	\sigma^2(N)&=\ln\left(\frac{1}{\E{Y}} - \frac{1}{N} + \frac{1}{P^q}\right) \nonumber\\
	&\approx \ln\left(\frac{1}{\E{Y}} + \frac{N^q}{\E{Y}^q}\right) \nonumber\\
	&= \ln\left(\frac{N^q}{\E{Y}^q}\right)+\ln\left(\frac{P^{q-1}}{N} + 1\right) \nonumber\\
	&\approx \ln\left(\frac{N^q}{\E{Y}^q}\right)+ \frac{P^{q-1}}{N}\quad \text{for $P^{q-1}/N\ll 1$} \nonumber\\
	&\approx q^2\ln\left(\e^{M(1-r)}\right)+ \frac{P^{q-1}}{N} \nonumber\\
	&= q^2M(1-r)+\frac{\e^{-M(1-r)q(q-1)}}{N}.
\end{align}

\begin{figure}
		\centering
			\includegraphics[width=0.5\linewidth]{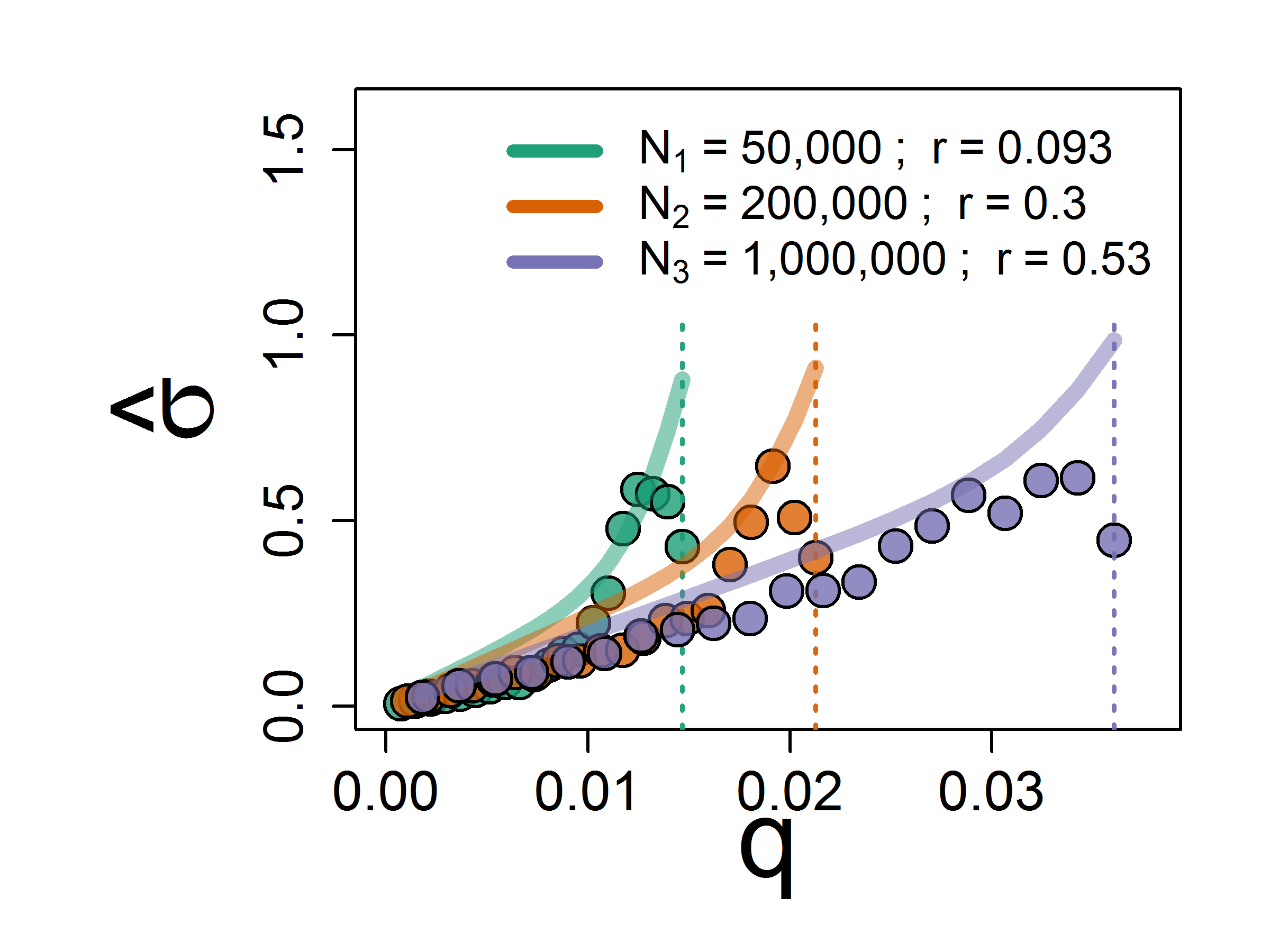}
		\caption{The theory predicts an approximately linear relationship between the standard deviation of $\ln(Y)$ and $q$. The three colors represent the relation for different population size levels. Interestingly, the fact that the curves are very close for most of the valid values of $q$ (i.e., $q\leq q_{\max}$, where $q_{\max}$ is the maximum $q$ for which $\E{Y|r,q,M,N}\geq 1$, shown as dotted vertical lines) means that $\sigma$ is a weakly varying function of population size $N$, as was noted in \cite{GomezLievanoYounBettencourt2012}. (The parameters for these curves are $M=808$, $a=-1.4827$ and $b=0.1456$.}
\label{fig:sigma2}
\end{figure}

Our model therefore predicts $\sigma(N)$ to be an approximately linear function of $q$, for a wide range of parameter values (see Fig.\ref{fig:sigma2}). Specifically,
\begin{align}
	\sigma(N)&\approx \sqrt{M(1-r)q^2 + \e^{-(1-r)M(q^2-q)}/N} ,\nonumber\\
	&\approx \sqrt{M(1-r)}q\quad\quad\text{for $P^{q-1}/N\ll 1$}.\label{eq:sigma}
\end{align}

To simplify matters, we average across population size, and we denote this measure of variance as $\sigma^2\equiv\langle \sigma^2(N)\rangle$:
\begin{align}
	\sigma^2&\approx \langle M(1-r)q^2 \rangle ,\nonumber\\
	&\approx M(1-a-b\langle \ln N \rangle)q^2,\label{eq:sigmaav}
\end{align}
where we have already made use of the assumption that diversity is a logarithmic function of population size, $r=a+b\ln(N)$.

Figure \ref{fig:sigma2} plots curves of Eq. \ref{eq:sigma} on top of simulations for three values of $N$ (and their three corresponding values of $r$).

%

\subsection{Why Does Diversity Scale Logarithmically with Size?}
\label{sc:whyrlogN}
In the main text, based on models of cultural evolution, we have assumed that diversity, $M~r(N) = D(N)$, scales approximately as a logarithmic function of $N$, $D(N)\approx A+B\ln(N)$. 

The question of why diversity grows in the way it does is a question about the mechanisms that drive the appearance of novel activities in cities, and in a social group more generally. There is an extensive body of research about the origins of innovation (see \cite{HenrichBoyd2002,Henrich2004Tasmanian,Strumsky20151445,Mesoudi2011,YounEtAl2015Universality}). We do not aim to engage too deeply into this question since our theory does not depend on the precise underlying mechanisms behind the growth of diversity. This is a research question that requires further investigation, and we provide below some reasonable mechanisms to explain why factors accumulate logarithmically with population size.

The two mechanisms we present are (i) skill-biased social-learning with incomplete inference, and (ii) random sampling from an extreme value distribution. They differ mainly in that the first analyzes cultural accumulation as a process that occurs \emph{within} individuals (i.e., individuals learn from each other), whereas the second analyzes cultural accumulation as a process that occurs at the systemic level (i.e., cities accumulate \emph{different} factors as population size grows). Fundamentally, this is a difference between \emph{intensive} and \emph{extensive} growth; the first is a statement about how much, on average, individuals know (their individual stock of skills), while the second is a statement about how much the city knows collectively (how many skills, which differ qualitatively among them, there are in the population). 

Both mechanisms that we present, however, share the essential feature that there is a selection process among random variables. This selection transforms statistical distributions into one of three extreme values distributions: 
\begin{itemize}
	\item If the underlying distribution (of trait values across individuals in the first case, or of the frequency of different factors in the second case) is thin-tailed (e.g., normal, exponential, poisson, etc.), and there is a selection for maximum values, the distribution will converge asymptotically to a Gumbel distribution. 
	\item If the underlying distribution has tails that fall as a power-law, through selection the distribution will converge to a Fr\'{e}chet distribution.
	\item If the underlying distribution has a finite right endpoint, the convergence is towards a Weibull distribution.
\end{itemize}
Of these three limiting distributions, the Gumbel has the largest domain of attraction \cite{EmbrechtsEtAl2013}.

\subsubsection{Cultural Evolution}
The model proposed by Henrich \cite{HenrichBoyd2002,Henrich2004Tasmanian} assumes that cultural factors (e.g., tools, beliefs, behaviors, skills, etc.) accumulate through an evolutionary process whereby individuals selectively imitate the most successful individuals in the population (i.e., ``prestige-biased transmission''). Copying the characteristics of the most successful individual, however, is an inferential process that is incomplete. To model this, Henrich and Boyd incorporate two essential features about human inference: first, that it is noisy (there are copying errors), and second, that it is biased (the copy is on average worse). The effects of selective imitation, under noisy and biased inferences, on the accumulation of factors at the population level, can be statistically separated using Price Equation \cite{frank1998foundations,Frank2012PriceEq}. 

At an abstract level, let us assume that there is an inheritable characteristic $z$ (the mode of transmission does not need to be genetic), and that different values of $z$ have different fitness $w$ (i.e., number of offspring). The average value of the characteristic across the population of individuals, $i=1,\ldots,N$, is $\bar{z}=\sum_i z_i/N$. This average changes from generation to generation due to both selective and other non-selective forces. The Price Equation is essentially a statistical decomposition of the change in the average characteristic value from one generation to the next, $\Delta \bar{z} = \bar{z}'-\bar{z}$, into these two forces:
\begin{equation}
	\bar{w}\Delta \bar{z} = \Cov{w, z} + \E{w\Delta z},
\end{equation}
where the first term in the right-hand side measures the selection force, and the second term measures other forces.

If $z_i$ is the size of the cultural repertoire of factors that individual $i$ carries in his or her social life, one can assume $\bar{z}$ changes through a process of evolution as described by Henrich's model of prestige-biased transmission and incomplete inference. If $f_i=w_i/(N\bar{w})$ is the probability that other members of the population will copy the characteristics of individual $i$, then Price Equation yields
\begin{equation}
	\frac{1}{N}\Delta \bar{z} = \Cov{f, z} + \E{f\Delta z}.
\end{equation}

The specific assumptions are mathematically expressed as follows:
\begin{enumerate}
	\item Prestige-biased transmission:
	\begin{equation}
		f_i = 
			\begin{cases}
				1,\quad\text{if $i=h$}\\
				0,\quad\text{if $i\neq h$},
			\end{cases}
	\end{equation}
	where $h$ is the most prestigious individual whose success comes from having the largest cultural repertoire, $z_h=\max\{z_1,\ldots, z_N\}$. This is a strong assumption that states that everyone attempts to imitate the single most prestigious individual in the society (however, see the subsection below about the speed of convergence). 
	
	\item Both $f$ and $z$ are random variables. The distribution of $f$ is $p=\Pr\{f=f_h\}$ and $1-p=\Pr\{f\neq f_h\}$  ($p=1/N$ if there is only one single prestigious individual). The distribution of $z$ is a Gumbel, $z\sim G(u, B)$. Hence, $\E{z}=u + \epsilon B$, where $\epsilon\approx 0.5772$ is the Euler-Mascheroni constant. The maximum of a sample of random variables Gumbel distributed is also Gumbel, yielding $z_h\sim G(u + B\ln(N), B)$.
	
	\item The incomplete inference is modeled by a random variable representing the errors in inference of individual $i$, $\Delta z_i$, in his or her attempt to imitate $h$. The assumption is that $\Delta z_i \sim G(w,B)$. The noisy aspect of inference is captured by the dispersion parameter $B$, and the downward bias is captured by $w$, plust the fact that the mode of the Gumbel is less than its mean.
	
\end{enumerate}

Assumptions 1 and 2 together imply that the first term in Price Equation is
\begin{align}
	\Cov{f,z} &= \E{fz} - \E{f}\E{z}, \nonumber\\
	&= (\E{fz|f=f_h} p + \E{fz|f\neq f_h} (1-p)) - \E{f}\E{z}, \nonumber\\
	&=\E{z_h} p - p\E{z}, \nonumber\\
	&=(u + B\ln(N) + \epsilon B)p - p(u + \epsilon B), \nonumber\\
	&=p B\ln(N).
\end{align}

Assumption 1, 2, and 3 together imply that the second term in Price Equation is
\begin{align}
	\E{f\Delta z} &= \E{f\Delta z|f=f_h} p + \E{f\Delta z|f\neq f_h} (1-p),\nonumber\\
	&= \E{\Delta z_h} p,\nonumber\\
	&= (w + \epsilon B)p.
\end{align}

All assumptions together yield
\begin{align}
	\frac{1}{N}\Delta \bar{z} = \frac{1}{N}\left(w + B(\epsilon + \ln(N))\right).
\end{align}

Finally, assuming that the total size of the cultural repertoire in the society is proportional to the cultural accumulation at the individual level, $D\approx \Delta\bar{z}$, one arrives at the relation 
\begin{equation}
	D \approx A + B\ln(N),
\end{equation}
where $A = w + B\epsilon$.

\subsubsection{Sampling from extreme value distributions}
We assume factors are sampled according to the population size of the city. Each factor $k$ has a different probability of being sampled, $f_k=\Pr\{K=k\}$, such that $\sum_k f_k = 1$. Suppose a city of population $N$ samples from this distribution $N$ times (imagine sampling from a bag of colors with replacement). The number of \emph{different values} of $k$ that the city draws is a function of population size, $D(N)$. Depending on the distribution $f_k$, the expected value $\mathrm{E}\{D(N)\}$ can take a different functional dependence on $N$.
\begin{figure}
		\centering
			\includegraphics[width=0.5\linewidth]{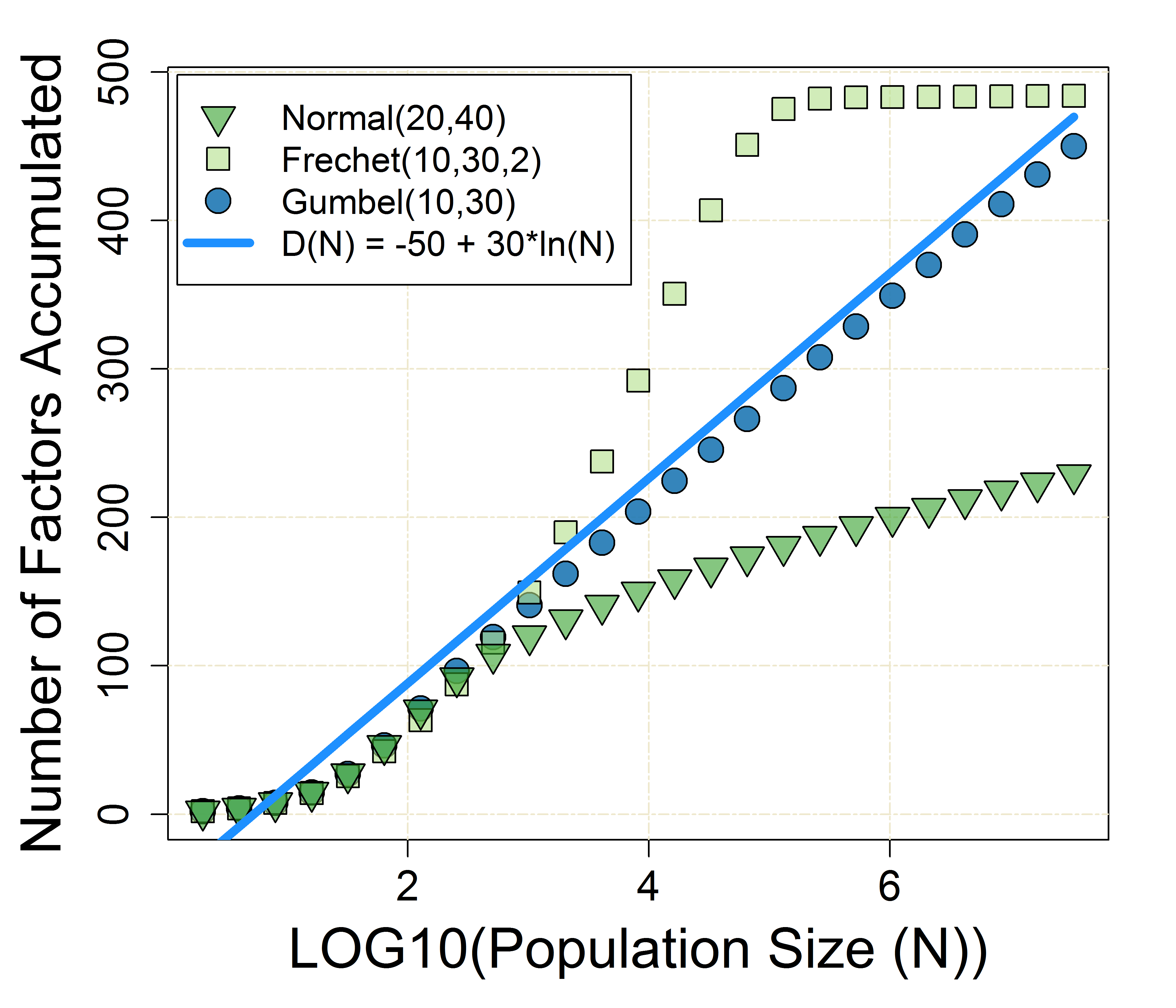}
		\caption{Demonstration of accumulation of factors when factors have different sampling probabilities. Gumbel distributions show a logarithmic accumulation with sample size $N$, in which the slope corresponds to the shape parameter.}
\label{fig:Gumbel}
\end{figure}

One can think of the values that $K$ can take as colors, and $f_k$ as the probability of sampling a given color. The question is thus how do the number of colors accumulate with sample size.

Let us represent the event of getting a new factor in the $N$th round of sampling, different from the factors a city already has, by $H_N=1$. Hence, the number of different factors (colors) in a population of $N$ is $D(N)=\sum_{i=1}^N H_i$. In other words, is the number of times a \emph{different color was sampled}. 

The probability that in the $N$th draw one samples a specific value $k$, new and different from the $N-1$ sampled before, is $(1-f_k)^{N-1}f_k$. As a result, $\Pr\{H_N=1\}=\sum_k (1-f_k)^{N-1}f_k$. This can also be interpreted as the probability of adding 1 to $D(N-1)$ in the $N$th draw. Hence, $\mathrm{E}\{D(N)\}=\sum_{i=1}^N \mathrm{E}\{H_i\}=\sum_{i=1}^N \sum_k (1-f_k)^{i-1}f_k$.

From the numerical simulations shown in Fig.\ref{fig:Gumbel} one can observe that if the distribution is Gumbel$(u,B)$, such that $f_k(u,B)\approx \frac{1}{B}\exp\left(-\frac{k-u}{B}\right)\exp\left(-\exp\left(-\frac{k-u}{B}\right)\right)$ (this is not a probability and it corresponds rather to the Gumbel density, but since we are calculating the probability for small intervals of size $\Delta k=1$, we make no distinction here), then $\mathrm{E}\{D_N\}\approx A + B\ln(N)$, for some constant $A$.

As already explained, the distribution with which cities sample would converge to a Gumbel if there is a selection process. Thus, for example, one can imagine that colors have an underlying (arbitrary) distribution, and cities sample several times, but only pick the maximum after several tries. This would amount to just sampling a single time from a Gumbel distribution.

\subsubsection{Convergence to the Gumbel}\label{sec:convergence}
How valid is the assumption of the Gumbel? If the convergence to this extreme value distribution is slow, our assumption about diversity being logarithmically related to population size may not be as general as we suppose. This can occur, for example, if in Henrich's model individuals learn from a few individuals only, e.g., from the most prestigious individual out of $K$ acquaintances instead of the whole population $N$. Picking the maximum from a small number of random variables may not be a selective force strong enough to drive the distribution to a Gumbel.

In the limit, however, the relation between diversity and population size would not change for $K<N$ in general. Results from Schl\"apfer et al. (2014) \cite{Schlapfer2014scaling} show that the average number of acquaintances in cities are well fitted by a nonlinear function of city size $K = f N^\alpha$, where $\alpha\approx 0.12$. Under the assumption that individuals have $K$ acquaintances to learn from, the process converges to the same result, since the relation to diversity will be $r = a' + b'\ln(K) = a' + b'\ln(f N^\alpha)$, which can be written as $r = a + b\ln(N)$, where $a = a' + b'\ln(f)$ and $b = \alpha b'$.

\begin{figure}[!ht]
		\centering
			\includegraphics[width=0.6\textwidth]{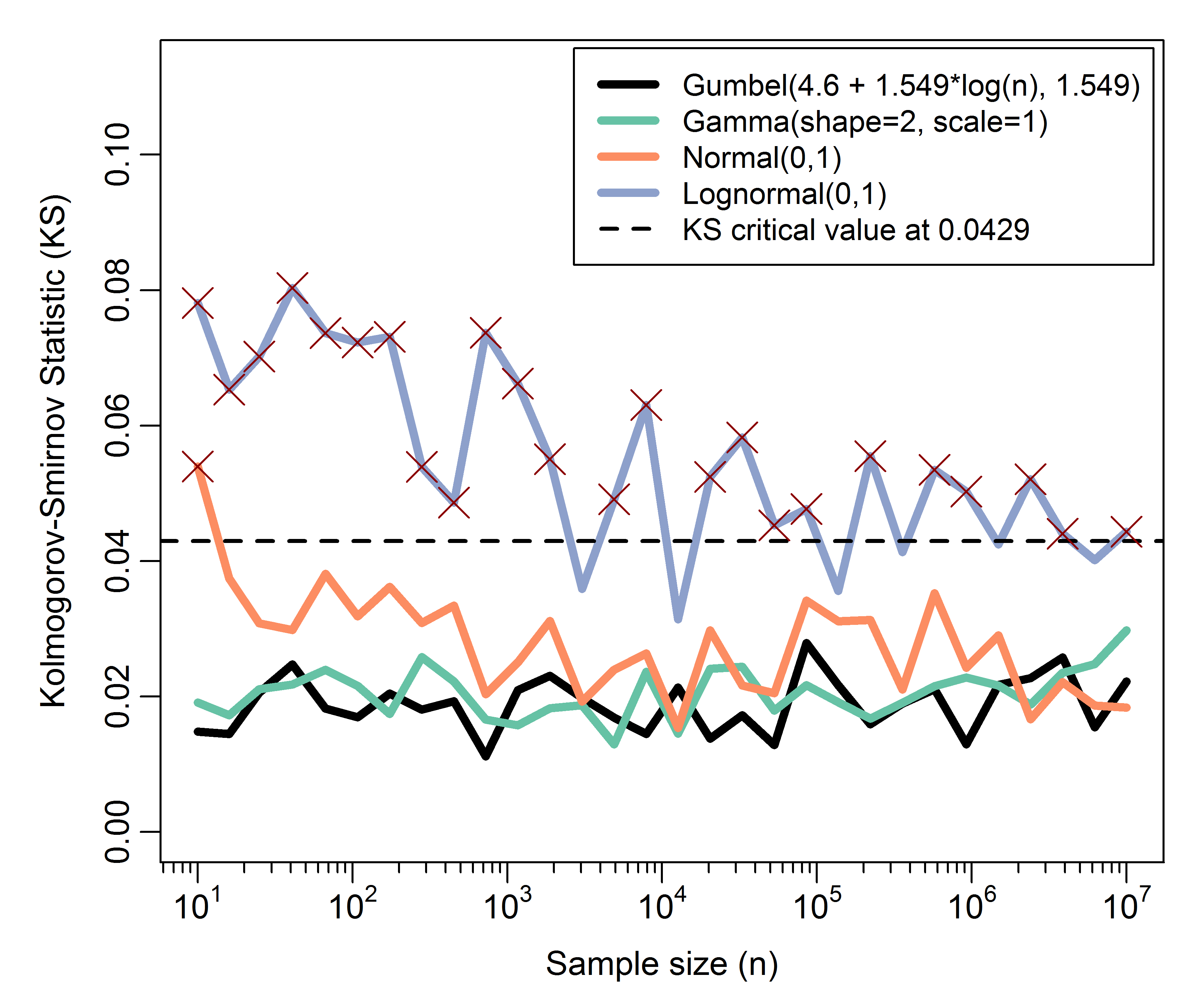}
		\caption{Monte Carlo simulations for choosing the maximum from a sample of size $n\in[10^1,10^7]$, and using 1000 simulations (i.e., 1000 i.i.d. maxima) to assess whether the distribution of these maxima approximates a Gumbel. Convergence of the maximum $z_h(n) = \max\{X_1, \ldots, X_n\}$ to a Gumbel distributed random variable as $n$ increases is depicted here as the Kolmogorov-Smirnov (KS) distance. On the vertical axis is the KS statistic and the horizontal axis is the sample size $n$ from which the maximum of one of the distributions, Gamma, Normal and Lognormal, is chosen. Below the dashed line (the critical value is computed using the formula $\sqrt{-0.5\ln(0.05/2)/numsims}$, where $numsims=1,000$, see \cite{Stephens1986tests}) one cannot reject the hypothesis that the random variables are distributed Gumbel, at a confidence level of 5\%.}
\label{fig:convergence}
\end{figure}

Still, the question is what size $n$ is large enough so that $z_h=\max_{i=1,\ldots,n}{z_i}$ is approximately Gumbel distributed. The convergence not only depends on $n$, but it also depends on the underlying distribution $z_i\sim\mathcal{D}$. Figure~\ref{fig:convergence} shows the distance between the distribution of 1,000 Monte Carlo Simulations $z_h$ and the Gumbel, measured by the Kolmogorov-Smirnov statistic, for different sample sizes $n$, and for three different distributions $\mathcal{D}\in\{Gamma, Normal, Lognormal, Gumbel\}$.

For the Gamma and Normal distributions, the convergence is fast and the distribution of their maxima is indistinguishable from a Gumbel. For the Lognormal, however, the convergence is very slow, and only maxima from populations above 10 million start to pass the KS test. The Lognormal distribution is special in the sense it lies in the frontier of the domain of attraction of the Gumbel \cite{EmbrechtsEtAl2013}. Hence, it is reasonable to assume that convergence is likely to be faster. Given that the Normal distribution also has a slow convergence to the Gumbel \cite{EmbrechtsEtAl2013}, the assumption of the Gumbel is reasonable even if individuals learn from a few number of acquantainces. However, further research is needed to discard the Lognormal distribution as a reasonable underlying distribution of the frequency of cultural traits.

\subsection{The pivot point of scaling}
\label{sc:pivotpoint}
Notice that in the model we have assumed there is a maximum number of factors $M$ into which a city can diversify. That is why $D(N)/M$ is a bounded number between 0 and 1 (where $a\equiv A/M$ and $b\equiv B/M$), which we treat as a probability $r$ that the city offers a factor or not. Notice that the population $N^*$ in which a city attains maximum diversification is such that $1=a+b\ln(N^*)$. Thus, $N^*=\mathrm{e}^{(1-a)/b}=\mathrm{e}^{1/s_1}$. The population $N^*$ is therefore the exponential of the inverse of the coefficient $s_1$ which relates the scaling exponent with the general prevalence of a given phenomenon. As argued in the text, the data suggests the coefficient $s_1$ is the same across urban phenomena, suggesting the population for which cities attain maximum diversification is, in some way, universal. According to our estimations, $N^*\approx 1.8\times 10^{14}$, a huge number.

Another way of understanding the meaning of $N^*$ is as a ``pivot point''.

Suppose a given linear relationship $y=mx+b$ between variable $y$ and $x$, and suppose this equation is conditioned on always passing through a particular point $(x^*,y^*)$. Let us call this point the ``pivot point''. It is easy to show, then, that the slope $m$ and the intercept $b$ are linearly related through the relation 
\begin{equation}
	m=(y^*/x^*)-(1/x^*)b. 
\end{equation}

Let us re-write once again the relationship our model predicts between the exponent and the baseline of urban scaling equations:
\begin{equation}
	\beta = 1-s_1\ln(Y_0).
\end{equation}

By comparing both equations we conclude that the pivot point implied by our model are
\begin{align}
	x^*&=y^*=1/s_1.
\end{align}
Recall from the previous section that $\ln(N^*)=1/s_1$. Hence, the scaling lines of urban phenomena, represented by the relation $y=Y_0~n^\beta$ are lines that pivot around the point $(\ln(N^*), \ln(N^*))$, in the log-log plane. In other words, urban scaling relationships across phenomena all pivot around the point of maximum diversification.

It is worth recalling that random noise in the relation $y=mx + b$ will create an artificial correlation between $\widehat{m}$ and $\widehat{b}$. If the noise is unbiased, then the regression lines should pass through the averages, and thus, the pivot point is $(x^*,y^*) = (\bar{x},\bar{y})$. Hence, the farther an estimated pivot point is from the averages $(\bar{x}, \bar{y})$, the less likely it is that the relationship between $m$ and $b$ is not a statistical artifact arising from the statistical correlation between $\widehat{m}$ and $\widehat{b}$.

\subsection{Geometrical Explanation of the Prediction Procedure}
There are an infinite number of lines that go through a single point. In the main text, however, we propose a prediction procedure to estimate the scaling line which only requires knowledge of a single data point. How is this possible?

The explanation is, of course, that we use two points. The first is the data point $(\ln(n_{city}), \ln(y_{city}))$, and the second is the pivot point of diversification $(\ln(N^*), \ln(N^*))$ (see section above about the pivot point).

What is interesting is that the pivot point can be estimated from observing only a single phenomenon. First, one estimates $\beta$, $\ln(Y_0)$, and $\sigma$. Second, one then solves for $G$ $H$ and $q$ (although $q$ is, in fact, not necessary):
\begin{align}
	G&=\frac{(\ln(Y_0))^2+(\beta-1)\ln(Y_0)\langle \ln(N)\rangle}{\sigma^2},\\
	H&=\frac{-(\beta-1)\ln(Y_0)-(\beta-1)^2\langle \ln(N)\rangle}{\sigma^2},\\
	q&=\frac{\sigma^2}{-\ln(Y_0)-(\beta-1)\langle \ln(N)\rangle}.
\end{align} 
And third, one computes $\ln(N^*)=G/H$. If this pivot point is the same for all urban phenomena, as our data suggests, then knowing a single data point of a specific phenomenon in an average city allows one to predict how this phenomenon will scale.

\section{Supplementary Figures}
\subsection{Linear relations in Figure 2 of the main text for different number of outliers}
We have implemented a kernel density estimation which allowed us to identify, in a principled way, outliers from a linear trend. Here we sequentially leave out outliers (i.e., points with the lowest estimated density), and we perform a linear regression over the rest of points. The aim of this exercise is to convince ourselves that the linear relationships indicated in Figure 2 that we show in the main text are robust. The coefficients of these regressions as reported here found to be reasonably stable. See below the Supplementary Figures 5-7.

\newpage
\begin{figure}[!ht]
		\foreach \x in {01,02,03}{
			\includegraphics[width=0.5\linewidth,left]{MultMech_rrr340_fig_HvsG_HvsGHlnN_outliers\x.pdf}
		}
		\includegraphics[width=0.3\linewidth]{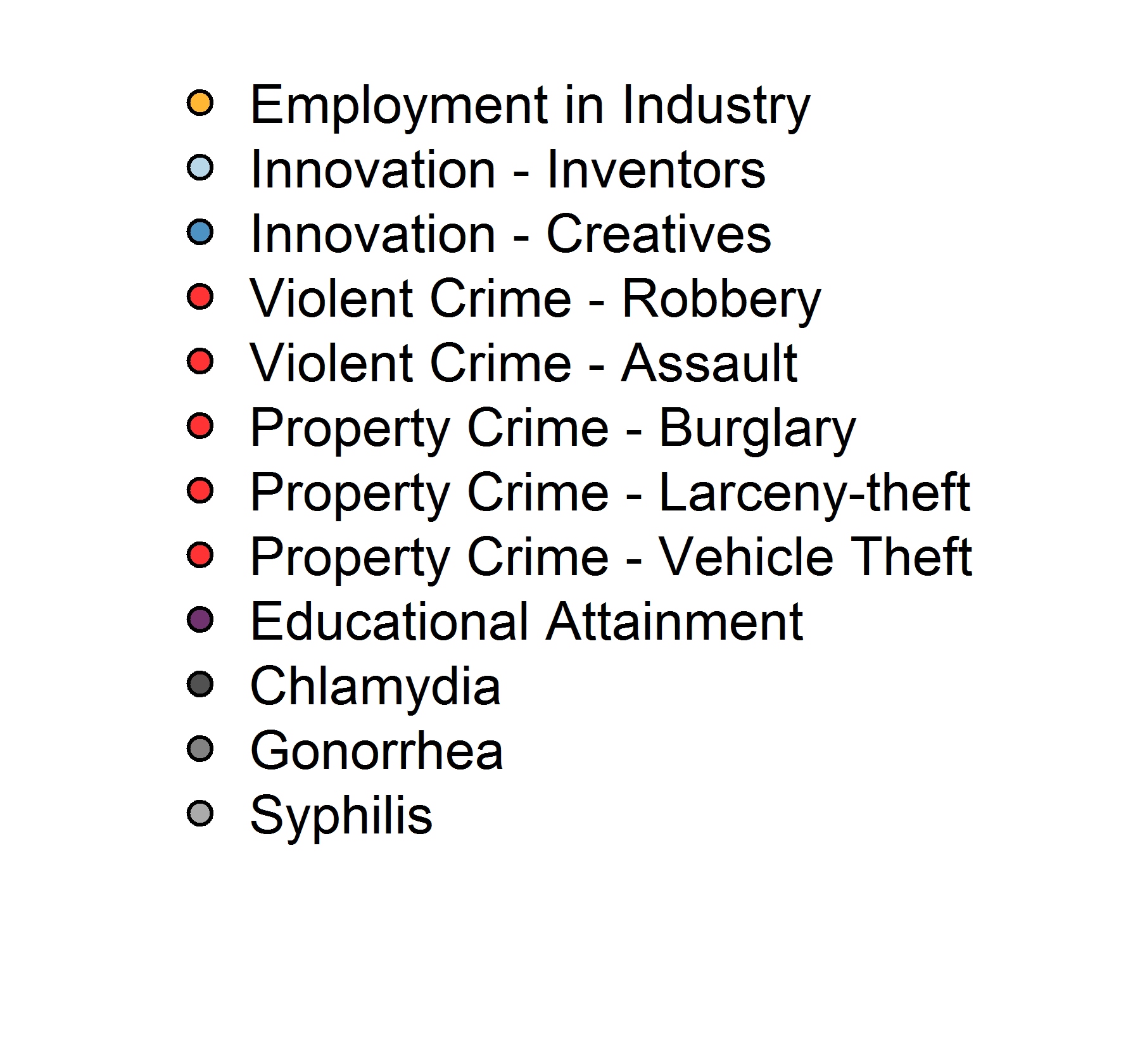}
		\caption{}
\label{fig1x}
\end{figure}
\newpage

\begin{figure}[!ht]
		\foreach \x in {04,05,06}{
			\includegraphics[width=0.5\linewidth,left]{MultMech_rrr340_fig_HvsG_HvsGHlnN_outliers\x.pdf}
		}
		\includegraphics[width=0.3\linewidth]{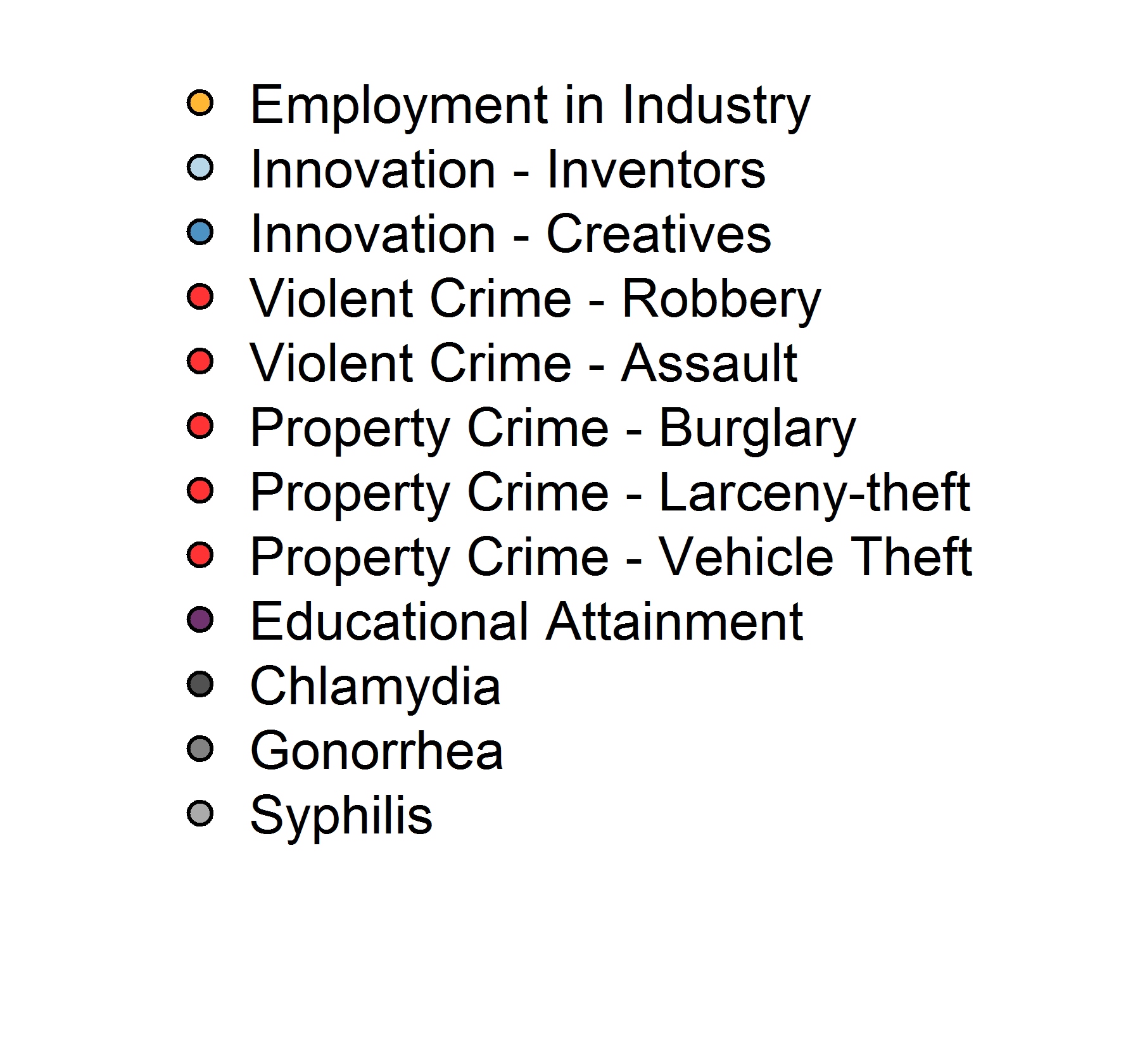}
		\caption{}
\label{fig2x}
\end{figure}
\newpage

\begin{figure}[!ht]
		\foreach \x in {07,08,09}{
			\includegraphics[width=0.5\linewidth,left]{MultMech_rrr340_fig_HvsG_HvsGHlnN_outliers\x.pdf}
		}
		\includegraphics[width=0.3\linewidth]{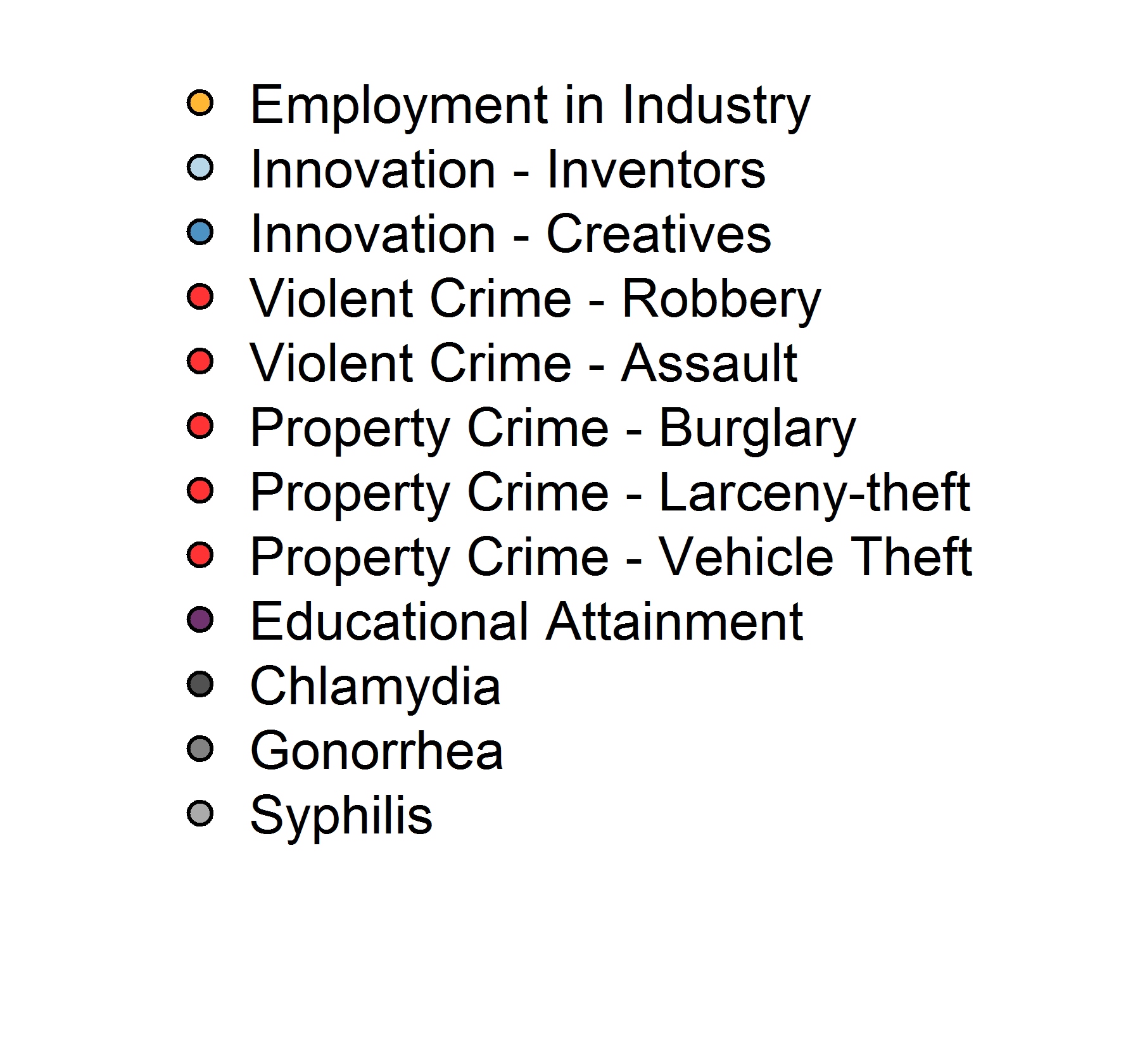}
		\caption{}
\end{figure}
\newpage

\section{Supplementary Data}
We provide a ZIP file with the data called ``Supplementary Data.zip'' that contains a single file for each urban phenomenon we studied (except for Sexually Transmitted Diseases that we kept in a single file), a README file, and a file ``ListUrbanPhenomena.xlsx'', which lists the different phenomena we used in our analysis with other parameters and field descriptions.



\end{document}